\documentstyle [eqsecnum,preprint,aps,epsf]{revtex}

\makeatletter
\def\footnotesize{\@setsize\footnotesize{10.0pt}\xpt\@xpt
\abovedisplayskip 10\p@ plus2\p@ minus5\p@
\belowdisplayskip \abovedisplayskip
\abovedisplayshortskip  \z@ plus3\p@
\belowdisplayshortskip  6\p@ plus3\p@ minus3\p@
\def\@listi{\leftmargin\leftmargini
\topsep 6\p@ plus2\p@ minus2\p@\parsep 3\p@ plus2\p@ minus\p@
\itemsep \parsep}}
\long\def\@makefntext#1{\parindent 5pt\hsize\columnwidth\parskip0pt\relax
\def\strut{\vrule width0pt height0pt depth1.75pt\relax}%
$\m@th^{\@thefnmark}$#1}
\long\def\@makecaption#1#2{%
\setbox\@testboxa\hbox{\outertabfalse %
\reset@font\footnotesize\rm#1\penalty10000\hskip.5em plus.2em\ignorespaces#2}%
\setbox\@testboxb\vbox{\hsize\@capwidth
\ifdim\wd\@testboxa<\hsize %
\hbox to\hsize{\hfil\box\@testboxa\hfil}%
\else %
\footnotesize
\parindent \ifpreprintsty 1.5em \else 1em \fi
\unhbox\@testboxa\par
\fi
}%
\box\@testboxb
} %
\global\firstfigfalse
\global\firsttabfalse
\def\tabular{\let\@halignto\@empty\@tabular}
\def\endtabular{\crcr\egroup\egroup $\egroup}
\expandafter \def\csname tabular*\endcsname #1{\def\@halignto{to#1}\@tabular}
\expandafter \let \csname endtabular*\endcsname = \endtabular
\def\@tabular{\leavevmode \hbox \bgroup $\let\@acol\@tabacol
   \let\@classz\@tabclassz
   \let\@classiv\@tabclassiv \let\\\@tabularcr\@tabarray}
\def\endtable{%
\global\tableonfalse\global\outertabfalse
{\let\protect\relax\small\vskip2pt\@tablenotes\par}\xdef\@tablenotes{}%
\egroup
}%
\def\lsim{ \,\, \vcenter{\hbox{$\buildrel{\displaystyle <}\over\sim$}} \,\,}
\def\gsim{ \,\, \vcenter{\hbox{$\buildrel{\displaystyle >}\over\sim$}} \,\,}
\makeatother

\def\c{_{\rm c}}

\def\Tc{T\c}
\def\mc{m\c}
\def\phic{\phi\c}
\def\eps{\epsilon}
\def\etal{{\it et al.}}
\def\vev{\langle\phi\rangle}
\def\mh{m_{\rm h}}
\def\mw{M_{\rm W}}
\def\mt{m_{\rm t}}
\def\meff{m_{\rm eff}}
\def\d{{\rm d}}
\def\SE{{\cal S}_{\rm E}}
\def\LE{{\cal L}_{\rm E}}
\def\Nc{N_{\rm c}}
\def\Piv{\Pi_{\rm v}}
\def\MSbar{$\overline{\rm MS}$}
\def\gammaE{{\gamma_{\rm\scriptscriptstyle E}}}
\def\qa{q_{\rm a}}
\def\qb{q_{\rm b}}
\def\xa{x_{\rm a}}
\def\xb{x_{\rm b}}
\def\sa{s_{\rm a}}
\def\sb{s_{\rm b}}
\def\ng{n_{\rm g}}
\def\Veff{V_{\rm eff}}
\def\GammaN{\Gamma_{\rm N}}
\def\sym{_{\rm sym}}
\def\asym{_{\rm asym}}
\def\ltwo{n}

\begin {document}

\preprint {UW/PT-93-24}

\title  {The $\epsilon$-expansion and the electroweak phase transition}

\author {Peter Arnold and Laurence G.~Yaffe}

\address
    {%
    Department of Physics,
    University of Washington,
    Seattle, Washington 98195
    }%
\date {November 1993}

\maketitle
\vskip -20pt

\begin {abstract}%
{%
{%
\advance\leftskip  -2pt
\advance\rightskip -2pt
Standard perturbative (or mean field theory) techniques are not adequate for
studying the finite-temperature electroweak phase transition in some cases of
interest to scenarios for electroweak baryogenesis.  We instead study the
properties of this transition using the renormalization group and the
$\epsilon$-expansion.  This expansion, based on dimensional continuation from
3 to $4{-}\epsilon$ spatial dimensions, provides a systematic approximation for
computing the effects of (near)-critical fluctuations. The $\epsilon$-expansion
is known to predict a first-order transition in Higgs theories, even for heavy
Higgs boson masses.  The validity of this conclusion in the standard model is
examined in detail.  A variety of physical quantities are computed at leading
and next-to-leading order in $\epsilon$.  For moderately light Higgs masses
(below 100~GeV), the $\epsilon$-expansion suggests that the transition is more
strongly first order than is predicted by the conventional analysis based on
the one-loop (ring-improved) effective potential.  Nevertheless, the rate of
baryon non-conservation after the transition is found to be {\em larger\/}
than that given by the one-loop effective potential calculation.  Detailed
next-to-leading order calculations of some sample quantities suggests that the
$\eps$-expansion is reasonably well behaved for Higgs masses below 100--200
GeV.  We also compare the $\eps$-expansion with large-$N$ results (where $N$
is the number of scalar fields) and find that the $\eps$-expansion is less
well behaved in this limit.

}%
\ifpreprintsty
\thispagestyle {empty}
\newpage
\thispagestyle {empty}
\vbox to \vsize
    {%
    \vfill \baselineskip .28cm \par \font\tinyrm=cmr7 \tinyrm \noindent
    \narrower
    This report was prepared as an account of work sponsored by the
    United States Government.
    Neither the United States nor the United States Department of Energy,
    nor any of their employees, nor any of their contractors,
    subcontractors, or their employees, makes any warranty,
    express or implied, or assumes any legal liability or
    responsibility for the product or process disclosed,
    or represents that its use would not infringe privately-owned rights.
    By acceptance of this article, the publisher and/or recipient
    acknowledges the U.S.~Government's right to retain a non-exclusive,
    royalty-free license in and to any copyright covering this paper.%
    }%
\fi
}%
\end {abstract}

\section {Introduction}

    The electroweak phase transition has received considerable
attention in the past few years because of its role in electroweak
scenarios for baryogenesis
\cite {baryogenesis}.
Important problems include the determination of
the order and strength of the phase transition
as a function of the Higgs boson mass,
and the calculation of the rate of baryon non-conservation near the transition.
In most previous work, the principle tool for studying these problems has
been the one-loop, ring-improved, finite-temperature effective potential for
the Higgs field.%
\footnote
    {%
    ``Ring-improved'' refers to the inclusion of lowest-order
    Debye screening corrections in the bare gauge and Higgs field propagators.
    }
As discussed below, the use of the ring-improved
loop expansion appears untrustworthy in many applications,
including the Higgs mass bounds for
baryogenesis in the minimal standard model
\cite {Dine,Shaposhnikov}.
The failure of the loop expansion is
exactly analogous to the failure of mean field theory
for describing critical behavior in condensed matter systems.
One alternative method for studying critical systems,
which has had considerable success in a variety of theories,
is to use the $\epsilon$-expansion to improve the
organization of the perturbation series.
The $\epsilon$-expansion generalizes the three spatial
dimensions of the theory to $4{-}\epsilon$ dimensions,
solves the theory when $\epsilon$ is small,
and then extrapolates to $\epsilon=1$
\cite {Wilson}.
In this paper,
we shall apply the $\epsilon$-expansion to the electroweak phase transition
with the goal of estimating various parameters of the phase transition.
We compute a number of observables at or near the transition
including the scalar correlation length, latent heat,
surface tension, free energy difference,
bubble nucleation rate,
and the sphaleron transition (or baryon non-conservation) rate.
Next-to-leading order results are presented for the correlation length
and the latent heat.
We discuss in detail the expected reliability of the $\epsilon$-expansion
results by (i) comparing leading and next-to-leading order contributions,
(ii) comparing results of the $\epsilon$-expansion with standard
perturbation theory in the limit of a small Higgs mass (where the
ordinary loop expansion remains reliable),
and (iii)
comparing $\eps$-expansion results when the number of scalar
fields is large to direct large-$N$ calculations.

In the remainder of the introduction, we review the need
for detailed knowledge of the electroweak phase transition in order
to determine the viability of electroweak baryogenesis scenarios,
and discuss the reliability of the standard loop expansion.
Then we briefly sketch the rather successful $\epsilon$-expansion results
for critical behavior in simple $\phi^4$ theories,
present an overview of the application of the $\epsilon$-expansion
to the standard electroweak theory,
and discuss some possible pitfalls of this approach.
We also review the existence of a non-perturbative magnetic mass in
SU(2) theory and discuss its implications for the $\epsilon$-expansion.
Detailed calculations follow in subsequent sections, which will be
outlined later in this introduction.

\subsection {Motivation}

    One of Sakharov's three famous requirements for baryogenesis
is that the universe be out of equilibrium.
In electroweak scenarios, baryogenesis occurs during the
electroweak phase transition and Sakharov's condition is met
if that transition is first order rather than second order.
First-order transitions proceed by the nucleation, expansion,
and coalescence of bubbles of the new phase inside the old---%
a highly non-equilibrium process.
However, the phase transition must not merely be first order,
it must be sufficiently {\it strongly\/} first order.
In particular, one constraint arises from considering
the rate of baryon number violation at the end of
the phase transition, when baryogenesis is completed.
This rate is exponentially sensitive to the expectation
value $\vev$ of the Higgs field responsible for symmetry breaking,
behaving as $\Gamma_{\rm B} \sim \exp\{-({\rm const.})\beta\vev/g\}$,
where $\beta$ is the inverse temperature and $g$ the electroweak coupling.
If $\vev$ is too small after the phase transition, then baryon-number
violation continues unabated after baryogenesis and the universe
relaxes back to an equilibrium state of zero net baryon number.
Any successful scenario requires that $\vev$ be sufficiently large
at the end of the transition so that baryon number violation is
effectively shut off.
Thus, one important goal of studies of the electroweak phase transition
has been to determine the strength of the transition,
as measured by the size of the discontinuity in $\vev$ across the transition.

    These issues have been previously studied using the one-loop
ring-improved finite-temperature potential for the Higgs field $\phi$.%
\footnote
    {%
    There have been several attempts to study corrections
    to these results using ``super-daisy'' or similar techniques based
    on solving mass-gap equations with partial inclusion of higher order terms.
    For a criticism of such methods,
    see Appendix~\ref{potential appendix} of Ref.~\cite {Arnold&Espinosa}
    and the discussion of overlapping momenta in Ref.~\cite {Boyd}.
    }
With this approximation to the potential,
one finds a first-order phase transition and
the Higgs expectation value immediately after
the transition can easily be evaluated.
In particular, one may study the strength of the transition
in the simplest interesting model:
the minimal standard model with a single Higgs doublet.
This is presumably only a toy model because,
unlike multiple-Higgs models,
it is unclear whether the minimal model incorporates
sufficient CP violation for baryogenesis.%
\footnote
    {%
    For the optimistic assessment, see Ref.~\cite {Farrar&Shaposhnikov}.
    }
Nevertheless, because of its small set of unknown parameters,
the minimal model provides a relatively uncluttered testing ground
for techniques used to analyze the phase transition.
In this model, employing the one-loop ring-improved potential,
Dine \etal\cite {Dine}
found that baryon number violation is turned off
at the end of baryogenesis only if the zero-temperature Higgs
mass $\mh(0)$ is less than 35 to 40~GeV.
A successful scenario for electroweak baryogenesis is then
incompatible with the experimental limit $\mh(0) > 60$~GeV.

\begin {figure}
\vbox
    {%
    \begin {center}
	\leavevmode
	
	\epsfbox [150 250 500 500] {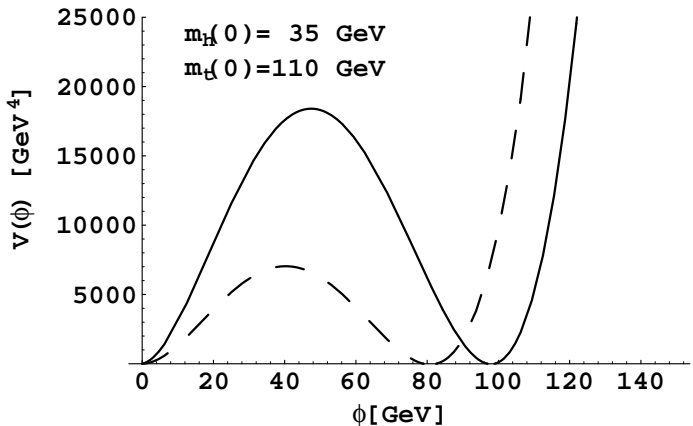}
    \end {center}
    \caption
	{%
	\label {figa}
	The effective potential at the critical temperature for
	$\mh(0)$ = 35 GeV and $\mt(0)$ = 110 GeV.
	The dashed and solid lines
	are the one-loop and two-loop results respectively.
	}%
    }%
\end {figure}

    This disaster is avoidable in multiple-Higgs models,
but in any model it will be important to understand the reliability of
predictions based on the one-loop potential.
So let us stick with the minimal model and review the reliability
of the above constraints.
The loop expansion parameter at high temperature is not simply
$g^2$ but rather $g^2 T/M$, where the characteristic mass scale
is set by the $W$-boson mass (at temperature $T$),
$M \sim g\phi$.
Computing $M$ at the phase transition, one finds that the
loop expansion parameter $g^2 T/M$ is of order $\lambda/g^2$.
(A review of this power counting may be found in section 2 of
Ref.~\cite {Arnold&Espinosa}.)
Hence, the loop expansion breaks down, and calculations
based on the one-loop potential are unreliable, unless
$\lambda \ll g^2$;
that is, unless the Higgs mass $\mh(0)$ at
zero temperature is much less than the $W$-mass $\mw(0)$.%
\footnote
    {%
    With our conventions $\mw^2(0) = g^2 \vev^2 / 4$
    and $\mh^2(0) = \lambda \vev^2 / 3$,
    so that $\lambda / g^2 = 3 \mh^2(0) / 4 \mw^2(0)$.
    }
This criteria is not obviously satisfied by the
$\mh(0)$ bound of 35 to 40 GeV found by Dine \etal,
and the importance of higher-order corrections will
be determined by all the detailed numerical factors
left out of this facile estimate of the loop expansion parameter.
To test the reliability of the one-loop results,
two-loop corrections to the potential were computed in
Refs.~\cite {Arnold&Espinosa,Bagnasco&Dine}
and are displayed at the critical temperature
in Fig.~\ref {figa} for $\mh(0) = 35$~GeV and a top quark mass
$\mt(0) = 110$~GeV.
Computed in Landau gauge,
$\vev$ increases by only 20\% when two-loop corrections are included.
But the height of the barrier in the potential increases by a factor of
almost three!
This dramatic change in the height suggests that the loop expansion
is unreliable.%
\footnote
    {%
    Ref.~\cite {Arnold&Espinosa} missed this signal of unreliability because,
    in numerical work, only the correction to $\vev$ was computed.
    }
One might optimistically hope that the correction
to the logarithm of the baryon violation rate $\Gamma_{\rm B}$
will be small because $\ln\Gamma_{\rm B}$ is proportional to $\vev$,
which has a relatively small two-loop correction.
However, this proportionality is only valid at leading order,
and $\ln\Gamma_{\rm B}$ has its own, independent
corrections controlled by $\lambda/g^2$.
There is no obvious reason why these corrections might
not be numerically large, similar to the correction to
the hump in the potential.

Because of these problems with the loop expansion,
it is clearly appropriate to explore other techniques,
such as the $\epsilon$-expansion, for studying
the electroweak phase transition.

\subsection {A success story: pure scalar theory.}%
\label {scalarsec}

    Before discussing the electroweak theory,
it will be useful to review briefly the $\epsilon$-expansion
in a much simpler model where it enjoys some success.
Consider a real scalar field theory
in $3+1$ spacetime dimensions at temperature $T$,
with the Euclidean Lagrangian
\begin {equation}
   \LE = {\textstyle {1\over2}} (\partial\phi)^2
       - {\textstyle {1\over2}}  \nu^2\phi^2
       + {\textstyle {1\over4!}} \lambda\phi^4 \,.
\label {Lising}
\end {equation}
One could attempt to analyze the phase transition by computing the
one-loop ring-improved potential.
With this approximation one would predict a first-order phase transition.
However, this computation also yields a loop expansion parameter
$\lambda T/\meff(T)$ of order 1 at the transition, and so
the one-loop result cannot be trusted.%
\footnote
    {%
    Again, see section 2 of ref.~\cite {Arnold&Espinosa}
    for a brief review in the same language used here.
    }
In fact, this model is in the same universality class as the Ising
model and is known to have a {\it second}-order phase transition.

    The $\epsilon$-expansion analysis of this transition
is well known and proceeds as follows.
The study of a second-order phase transition
requires exploring the infrared behavior of the theory.
In the Euclidean formulation of equilibrium finite-temperature
quantum field theory,
space-time is periodic in the time direction with period $\beta \equiv 1/T$.
For spatial distances $r$ large compared to $\beta$,
fluctuations in the Euclidean time direction are irrelevant
and the theory reduces to an effective three-dimensional theory.
Up to corrections suppressed by powers of $\beta/r$,
the only effect of decoupled temporal fluctuations is to renormalize
the parameters of the three dimensional theory,
{\em i.e.}, masses and coupling constants,
from their original four-dimensional values.
The relationship between the three- and four-dimensional parameters
is perturbatively calculable.
In the case at hand, one finds that the
mass parameters in the three- and four-dimensional theories are related by
\begin {equation}
   m^2 = - \nu^2 + {\textstyle {1\over24}} \, \lambda T^2
       + O(\lambda \nu^2, \lambda^2 T^2) \,.
\label {mrenorm}
\end {equation}
The second term is the standard one-loop thermal contribution to the mass and
is responsible for the phase transition.
By adjusting $T$, one can approach the phase transition
by tuning the mass $m$ of the three-dimensional theory toward zero.
The study of second-order (or weakly first-order) phase transitions
is therefore the study of the infrared behavior of 
three-dimensional theories.

\begin {figure}[b]
\vbox
    {%
    \begin {center}
	\leavevmode
	
	\epsfbox [72 390 520 410] {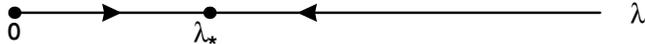}
    \end {center}
    \caption
	{%
	\label {figb}
	The renormalization group flow in $\lambda\phi^4$ theory.
	}%
    }%
\end {figure}

    Now consider
this theory in $d = 4{-}\epsilon$ spatial dimensions, instead of 3,
where $\epsilon$ is finite but small compared to 1.
The infrared behavior
can be studied by following the flow of relevant
couplings using the renormalization group (RG).
For a scalar theory, one has
\begin {equation}
   \mu {d\over d\mu} \lambda(\mu) = -\epsilon\lambda + \beta(\lambda).
\label {lamrun}
\end {equation}
The first term, which vanishes in the four-dimensional limit,
reflects the trivial classical scaling arising from the fact that
the interaction $\int d^d x \, \phi^4$ has engineering dimension
$-\epsilon$.
The function $\beta (\lambda)$ is the usual four-dimensional $\beta$-function
given perturbatively by%
\footnote
    {%
    Throughout this paper, we use a minimal subtraction renormalization scheme.
    As is well known, in this scheme the $\beta$-function has no additional
    $\epsilon$-dependence
    \cite {Gross}.
    }
\begin {equation}
   \beta(\lambda) = {3\over(4\pi)^2} \, \lambda^2 + O(\lambda^3) \,.
\label {betalam}
\end {equation}
The renormalization group flow is depicted qualitatively in Fig.~\ref {figb},
where the arrows denote flows into the infrared ({\em i.e.}, decreasing $\mu$).
There is an infrared-stable fixed point at
\begin {equation}
\label {fixpt}
   \lambda_* = {(4\pi)^2\! \over3} \, \epsilon + O(\epsilon^2) \,.
\end {equation}
Such a fixed point indicates a second-order phase transition since,
if the temperature has been adjusted so that $m$ is zero,
the behavior of the theory at the fixed point in $\lambda$ will then
look the same on all distance scales.

    Note that the fixed point coupling is small, $\lambda \ll 1$,
provided that $\epsilon \ll 1$.
For this reason, the one-loop result for
the $\beta$-function (\ref {betalam}) is sufficient to find the
fixed-point to leading order in $\epsilon$.
Renormalization-group improved perturbation theory in $\lambda$
is therefore equivalent to an expansion in powers of $\epsilon$.
It is for this reason that calculations of the phase transition
are more tractable in $4{-}\epsilon$ dimensions, when $\epsilon$ is small,
than in three dimensions.

    By computing anomalous dimensions of various operators
at the fixed point, one can extract the critical exponents
associated with the phase transition.
For instance, one finds in a three-loop calculation that
the susceptibility exponent $\gamma$, defined by $\chi \sim |T-\Tc|^\gamma$
where $\chi$ is the susceptibility, is given by
\cite {Wilson,Gorishny}
\begin {equation}
\label {Igamma}
   \gamma = 1 + 0.167 \, \epsilon + 0.077 \, \epsilon^2 - 0.049 \, \epsilon^3
   + O(\epsilon^4) \, .
\end {equation}

    Perturbation expansions are asymptotic in $\lambda$, and the terms
in the expansion start growing in magnitude at orders $n \gsim O(1/\lambda)$.
Expansions in $\epsilon$ are therefore also asymptotic,
with terms growing in magnitude at orders $n \gsim O(1/\epsilon)$.
What does this imply when one finally returns to the three-dimensional
theory by sending $\epsilon\to 1$?
If one is lucky, $O(1/\epsilon)$ really means something like three or four
when $\epsilon = 1$ and the first few terms of the series will be useful.
If one is unlucky, no terms in the expansion will be useful.
Whether or not one will be lucky cannot be determined in advance of an
actual calculation.
{}From the result (\ref {Igamma}), we see that the terms displayed do indeed
get smaller for $\epsilon=1$,
and their sum in 1.195.
This compares favorably with results determined by high-order Borel
resummation techniques\cite {Zinn-Justin}, which give
$1.2405 \pm 0.0015$.
(This value is consistent with numerical results using lattice
methods\cite{Gupta}.)
So the straightforward use of the $\eps$-expansion above appears to get within
4\% of the true answer---%
a stunning achievement.

    Another example, which isn't quite as nicely behaved, is the anomalous
dimension $\eta$ of the two-point correlation function,
defined by $\langle\phi(r)\phi(0)\rangle \sim r^{2-d-\eta}$.
Because its leading order contribution happens to vanish,
the exponent $\eta$ is a small number.
One finds \cite {Wilson,Chetyrkin}
\begin {equation}
\label {Ieta}
    \eta = 0.0185 \, \epsilon^2 + 0.0187 \, \epsilon^3
         - 0.0083 \, \epsilon^4 + 0.0359 \, \epsilon^5 + O(\epsilon^6) \, .
\end {equation}
If one takes $\epsilon \to 1$ and adds terms until they start
increasing, one obtains $\eta = 0.0289$.
The correct answer is believed to be
$\eta \approx 0.035$ \cite {Zinn-Justin,Gupta}.

\subsection {Electroweak theory}

    We now turn to the application of the $\epsilon$-expansion
to electroweak theory.%
\footnote
    {%
    A few authors have recently made partial attempts to apply
    $\eps$-expansion techniques to current problems in the study of
    the electroweak phase transition.  Ref.~\cite{Alford}\ examines
    related problems in the cubic anisotropy model, where first-order
    transitions were originally studied with the $\eps$-expansion
    by Rudnick~\cite{Rudnick}.
    Ref.~\cite{Rocky} gives a very rough, heuristic attempt to incorporate
    renormalization-group results into an order-of-magnitude
    estimate of corrections to the conventional one-loop potential.
    }
We shall ignore the Weinberg mixing angle,
and concentrate on an SU(2) gauge theory with a single complex Higgs doublet.
One must first reduce the theory to an effective three-dimensional one,
and then map out the renormalization group flows in $4{-}\epsilon$ dimensions.
This procedure has been analyzed by Ginsparg for general non-Abelian
gauge theories \cite {Ginsparg}.
When constructing the three-dimensional theory,
one finds that the $A_0$ component of the four-dimensional gauge
field picks up a Debye screening mass of order $g T$.
The three-dimensional theory then consists of the SU(2) Higgs doublet
$\Phi(\vec x)$, a three-dimensional SU(2) gauge field $\vec A(\vec x)$,
and a massive scalar $A_0(\vec x)$ that is in the adjoint
representation of SU(2).
We are interested in the infrared behavior of the theory,
however, and the massive $A_0$ field will decouple from distance
scales large compared to $1/gT$.
One may integrate out its effects as well,
leaving a three-dimensional SU(2)-Higgs theory of $\Phi$ and $\vec A$.

\begin {figure}
\vbox
    {%
    \begin {center}
	\leavevmode
	
	\epsfbox [72 260 520 550] {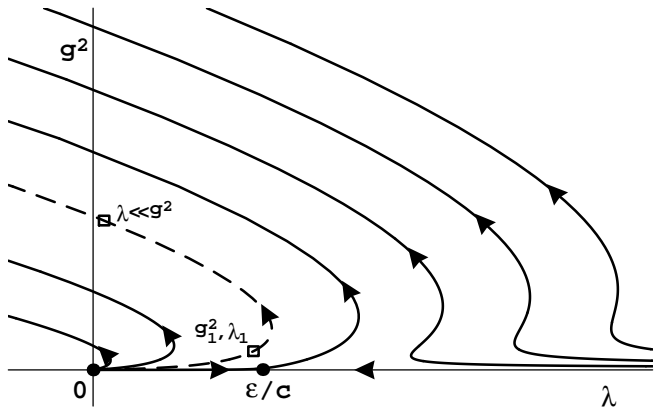}
    \end {center}
    \caption
	{%
	\label {figc}
	The renormalization group flow for an SU(2)-Higgs theory.
	The dashed line is the trajectory which flows from the
	initial couplings $(g_1^2, \lambda_1)$ into the region where
	$\lambda \ll g^2$.
	}%
    }%
\end {figure}

Now consider instead a $4{-}\epsilon$ dimensional SU(2)-Higgs theory.
The one-loop renormalization group equations have the form
\begin {mathletters}
\label {surng}
\begin {eqnarray}
    \mu {d g^2\over d\mu} &=& - \epsilon \, g^2 + \hat\beta_0 \, g^4 ,
\\
    \mu {d\lambda\over d\mu} &=& - \epsilon \, \lambda +
                 (\hat a \, g^4 + \hat b \, g^2 \lambda + \hat c \, \lambda^2).
\end {eqnarray}
\end {mathletters}%
The precise numerical coefficients, as well as the explicit solution,
will be given in section~2.
One finds that the flows have the form shown in Fig.~\ref {figc}.
There is no infrared stable fixed point,
and all trajectories flow to the region $\lambda < 0$
where the Higgs sector appears to become unstable.
This instability, and the absence of stable fixed points,
suggest that the phase transition is always first-order
in $4{-}\epsilon$ dimensions
\cite {Ginsparg}.

    The last point deserves more careful analysis,
which can be provided by recalling the previous discussion
of the one-loop effective potential.
Notice that, although the couplings may start at a point $(g_1^2,\lambda_1)$
where the naive loop expansion parameter $\lambda/g^2$ is not small,
they eventually flow into the region $\lambda \ll g^2$.
At this point one can apply the familiar analysis based
on computing the effective potential.
As we shall see, the phase transition is indeed first-order
when $\epsilon$ is small and its properties can be reliably computed.

    In order to make contact with $\epsilon = 1$,
it is useful to consider how Fig.~\ref {figc} changes with $\epsilon$.
The one-loop renormalization group equations have the
interesting property that the trajectories in the ($g^2$,$\lambda$)
plane are {\it independent\/} of $\epsilon$ if $g^2$ and $\lambda$
are rescaled by $\epsilon$, but the rate at which those
trajectories are traversed is exponentially sensitive to $\epsilon$.
Specifically, by rewriting
\begin {equation}
   g^2 \equiv \epsilon \, \tilde g^2,
   \qquad
   \lambda \equiv \epsilon \, \tilde\lambda,
   \qquad
   \mu \equiv \tilde\mu^{1/\epsilon},
\label {grescale}
\end {equation}
the equations become independent of $\epsilon$,
\begin {mathletters}
\label {rngrescale}
\begin {eqnarray}
        \tilde\mu {d\tilde g^2\over d\tilde\mu}
    &=&
        - \tilde g^2 + \hat\beta_0 \, \tilde g^4 ,
\\
        \tilde\mu {d\tilde\lambda\over d\tilde\mu}
    &=&
        - \tilde \lambda
        + (\hat a \, \tilde g^4 +
           \hat b \, \tilde g^2 \tilde\lambda +
           \hat c \, \tilde\lambda^2) \,.
\end {eqnarray}
\end {mathletters}%
Hence, for these one-loop equations,
Fig.~\ref {figc} does not change with
$\epsilon$ provided we label the axes as $g^2/\epsilon$ and $\lambda/\epsilon$.

    We are now in a position to outline our basic approach for using the
$\epsilon$-expansion to compute quantities
associated with the phase transition.
This approach is closely related to the methods used years ago
by Chen, Lubensky, and Nelson to study U(1)-Higgs theory
({\em i.e.}, superconductivity) \cite {Chen&Lubensky&Nelson}
and before that by Rudnick to study the
cubic anisotropy model \cite {Rudnick}.%
\footnote{
   See part II, chapter 4 of Ref.~\cite {Amit} for a review.
}
Start by considering the original four dimensional theory and the
effective three dimensional theory, regulating both with dimensional
regularization and renormalizing at the scale $\mu_1=T$.
Relate the three dimensional couplings to
the four dimensional ones by computing perturbatively, in both theories,
physical quantities characterized by the momentum scale $T$ and equating
the results.
For example, one can compute the four-point scalar
amplitude when the external momenta are of order $T$.
One finds a perturbative relation between the three- and four-dimensional
couplings which is trivial at leading order:
\begin {mathletters}
\begin {eqnarray}
   g_1^2 &=& \bar g^2(T) + O(\bar g^3), \qquad
\\
   \lambda_1 &=& \bar\lambda(T) + O(\bar g^3, \bar \lambda^2).
\end {eqnarray}\label {ginit}%
\end {mathletters}%
Here,
$g_1^2$ and $\lambda_1$ are the three-dimensional couplings at a scale
$\mu_1=T$,
and $\bar g^2$ and $\bar\lambda$ are the couplings
of the original 4-dimensional theory.
The three dimensional couplings $g_1^2$ and $\lambda_1$ are the starting
points for the renormalization group flow depicted in Fig.~\ref {figc}.
Now go from three to $4{-}\epsilon$ dimensions while holding $g_1^2/\epsilon$
and $\lambda_1/\epsilon$ fixed,
so that the relative position in Fig.~\ref {figc} does not change.
Finally, use the renormalization group to flow to an equivalent theory
in the region $\lambda \ll g^2$.
The phase transition in this theory can now be studied using
a loop expansion to compute the effective potential.
Note that, for small $\epsilon$,
scaling the couplings $g_1^2$ and $\lambda_1$ to be $O(\epsilon)$
while holding $\lambda_1 / g_1^2$ fixed,
automatically implies that $g_1^4 \ll \lambda_1 \ll 1$.

    In this paper, we shall not concern ourselves with the calculational
details of the higher-order matching (\ref {ginit}) of the
three-dimensional couplings with the original four-dimensional ones.
We shall instead focus on the infrared behavior of the
three-dimensional theory as predicted by the $\epsilon$-expansion.
This is appropriate if the original four-dimensional couplings
$\bar g^2$ and $\bar \lambda$ are taken to be very small while the ratio
$\bar \lambda/\bar g^2$ is held fixed.
So we shall study the limit
\begin {equation}
   \label{g-assumptions}
   \bar g^2, \bar \lambda \rightarrow 0,
   \qquad \hbox{with} \qquad
   {\bar\lambda/\bar g^2} ~ \hbox{fixed}.
\label {coupling-inequalities}
\end {equation}
In practice, this means that we will assume that
$\bar\lambda \sim \bar g^2 \ll 1$ but $\lambda \gg g^3,~g^4,$
etc.
This is appropriate for the real electroweak theory
if the Higgs mass is small compared
to 1 TeV but not much smaller than the W-mass.
We make this assumption that the couplings are small only to
simplify the relationship between the four- and three-dimensional theories.
When working within the three-dimensional theory itself, however,
we shall generally present formulas that are valid even when
$g_1^2$ and $\lambda_1$ are not assumed small.

\subsection {Outline}

    Our presentation in section II begins with the one loop analysis.
The explicit solution of the one-loop renormalization group flow equations
is presented, followed by the evaluation of the one-loop effective potential
(in $4{-}\epsilon$ dimensions) when the scalar self-coupling vanishes.
This information is then used to compute, to one-loop order within the
$\epsilon$-expansion,
the scalar correlation length at the phase transition
in both the symmetric and asymmetric phases,
the baryon violation and bubble nucleation rates at the transition,
the latent heat of the transition, surface tension, and
the free energy difference between symmetric and asymmetric phases
near the transition.
The predictions of the $\epsilon$-expansion for these physical quantities
are then compared to the results obtained from one-loop perturbation
theory directly in three dimensions,
in the limit where $\lambda \ll g^2$.

    Section III contains two loop analysis.
The two-loop renormalization group equations are presented, along with
their explicit solutions,
followed by the two-loop effective potential for $\lambda=0$.
Two-loop corrections to the scalar correlation length and latent heat
are computed.
We again compare results with direct three-dimensional perturbation
theory in the $\lambda \ll g^2$ limit.
More significantly, by examining the relative size of the two-loop
corrections, we can test whether the $\eps$-expansion is well-behaved
in the case $\lambda \gsim g^2$ of moderate-to-heavy Higgs mass,
where conventional perturbation theory fails.

    Section IV examines the $\epsilon$-expansion when one increases
the number of scalar fields, $N$.
Near four dimensions, when the number of charged scalar fields
is sufficiently large ($N > \Nc$),
two additional renormalization group fixed points appear:
an infrared stable fixed point and a tricritical point.
We compute the first two terms in the $\eps$-expansion of
the critical value $\Nc$ at which these fixed points first appear.
We also examine the $N \to \infty$ limit of the slope $\lambda / g^2$
of the tricritical line separating theories with
first and second order transitions, and compare the $\eps$-expansion
result with that obtained directly in three-dimensions
using large-$N$ methods.

    The overall interpretation of our results is summarized in the conclusion.
Various details of two-loop calculations are relegated to appendices.

\subsection {Possible Misgivings}
\label {dasguptasec}

\begin {figure}
\vbox
    {%
    \begin {center}
	\leavevmode
	
	\epsfbox [150 260 500 540] {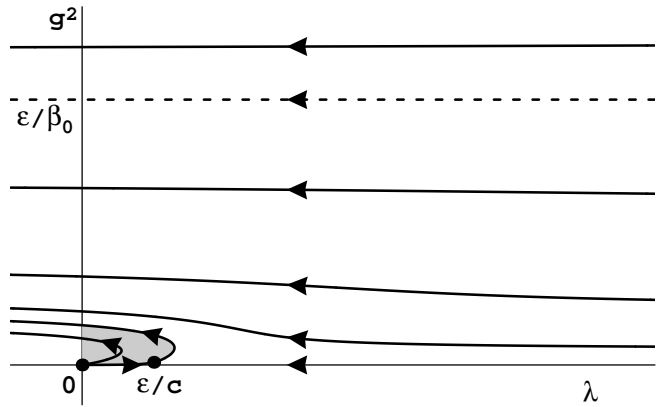}
    \end {center}
    \caption
	{%
	\label {figd}
	The renormalization group flow for a U(1)-Higgs theory with
	$N < \Nc$.
	}%
    }%
\end {figure}

    Before embarking on explicit calculations using the $\epsilon$-expansion,
it is worth considering whether the endeavor is doomed before it begins.
In particular, the $\epsilon$-expansion predicts that the phase transition is
always first order, even when $\lambda_1/g_1^2$ is arbitrarily large.
Some authors have dismissed this result based on the apparent failure
of analogous arguments for superconductivity.%
\footnote
    {%
    For a nice review of such arguments, see Ref.~\cite {March-Russel}.
    }
At large distances,
simple superconductors are described by Landau-Ginzburg theory,
which is just U(1) gauge theory with a single, charged, complex scalar field.
The renormalization group equations in $4{-}\epsilon$ dimensions have
the same form as Eq.~(\ref {surng}) with $\beta_0$ positive.
The flow is shown in Fig.~\ref {figd}.
As observed by Halperin \etal, \cite {Halperin},
the $\epsilon$-expansion predicts the superconducting phase transition
to be always first order, even deep in the Type II regime,
$\lambda_1 \gg g_1^2$.

The predicted strength of the transition is too small to measure in
superconductors, but detailed studies have been made of the critical
properties of smectic-A liquid crystals, which are argued to be
in the same universality class as superconductors \cite {liquid-crystals}.
The predicted first-order transition is often not observed.
Motivated by this discrepancy, Dasgupta and Halperin \cite {Dasgupta&Halperin}
made a lattice study of the superconducting transition.
They used a U(1) lattice gauge theory in three dimensions with a fixed-length
Higgs field; that is, their bare theory had $\lambda_1\to\infty$
with $m^2 / \lambda_1$ fixed.
Simulating a bare theory that had $g_1^2=5$,
they found no numerical evidence of a first-order phase transition.

Their numerical results are not ironclad.
The height of the specific heat peak was measured on lattices of size
$3^3$, $5^3$, $10^3$, and $15^3$.
In a first-order phase transition, the height should grow
like the volume ${\cal V}$ in the large-volume limit;
for a second-order transition, it should grow like ${\cal V}^{2\alpha/3}$
(or $\ln {\cal V}$ if the specific heat exponent $\alpha \approx 0$)
\cite {Dasgupta&Halperin}.
Results on these four lattice sizes did not support
a linear relationship between the height and ${\cal V}$.
However it is not clear if these simulations reached the large volume limit.
As a rough guide, one may consider the predictions of the
$\epsilon$-expansion.
If we use the one-loop flows and set $\epsilon=1$,
we find that the scale change necessary for the initial theory with
$\lambda_1 = \infty$ and $g_1^2 = 5$ to run to the region
$\lambda < 0$ is roughly 9.
(Detailed formulas are given in section 2.)
This suggests that the width of domain walls,
or the minimal size of critical bubbles
nucleated in a first-order phase transition,
will be of order $9$ lattice units.
Therefore, the linear growth of
the specific-heat peak with ${\cal V}$ need not begin until
the lattice size is larger than $O(9^3)$.
Given that the lattices of \cite {Dasgupta&Halperin} range in size from
$3^3$ to $15^3$, the reliability of the conclusions is clearly sensitive
to whether the condition ${\cal V} \gg O(9^3)$ really means something
like ${\cal V} \ge 4^3$ versus something like ${\cal V} \ge 18^3$.

    Nevertheless, given the other evidence from liquid crystals,
let us put this concern aside and hypothesize that the phase transition
studied in \cite {Dasgupta&Halperin} is in fact second order.
The presence of a continuous transition in a fixed-length U(1)-Higgs theory
would imply that the $\epsilon$-expansion does a terrible job describing
the flow of theories starting near $\lambda_1 = \infty$.
These theories must flow to some infrared-stable fixed
point not seen within the $\epsilon$-expansion.
However,
as noted after Eq.~(\ref {coupling-inequalities}),
in the application to electroweak theory
one is interested not in theories that start at strong coupling
with $\lambda_1=\infty$,
but in perturbative
theories that start near the unstable Gaussian fixed point at
$\lambda=g^2=0$.
The latter theories need not lie within the domain of attraction
of the putative fixed point,
and might instead be correctly described by the $\epsilon$-expansion.
Within the $\epsilon$-expansion, all theories which start near $\lambda=g^2=0$
stay inside the shaded region of Fig.~\ref {figd}.
Recall that at leading order the $\epsilon$-expansion is based on
renormalization-group improved one-loop perturbation theory.
One way to view the success of the $\epsilon$-expansion in pure
$\lambda\phi^4$ theory is to note that,
even though the fixed point coupling $\lambda$ is formally $O(1)$ when
$\epsilon\to 1$, the coupling is not such a large number
that higher-order corrections in $\lambda$ cause dramatic changes.
If we are similarly lucky,
the couplings in the shaded region of Fig.~\ref {figd} will not be so large
when $\epsilon\to 1$ as to render the $\epsilon$-expansion useless.

\begin {figure}
\vbox
    {%
    \begin {center}
	\leavevmode
	
	\epsfbox [150 250 500 530] {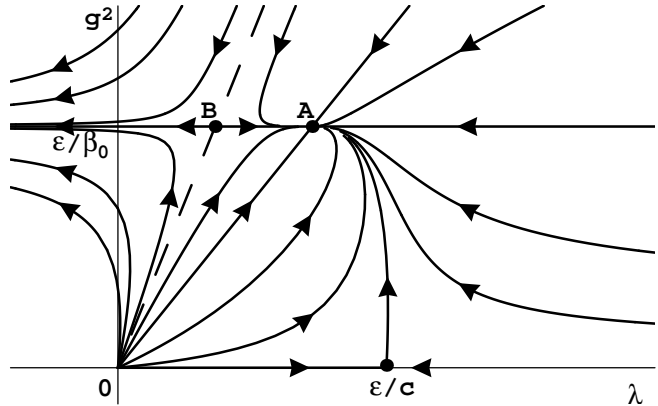}
    \end {center}
    \caption
	{%
	\label {fige}
	The renormalization group flow for a U(1)-Higgs theory with $N > \Nc$.
	Point $A$ is an infrared-stable fixed point,
	while $B$ is a tricritical point.
	The dashed line is the tricritical line which separates
	theories with first and second order transitions.
	}%
    }%
\end {figure}

An important general point
is that phase transitions are more likely to be second order in lower
dimensions due to the increased importance of long-distance fluctuations.
A prediction of a second-order phase transition in $4{-}\epsilon$ dimensions
may be more robust as $\epsilon\to 1$ than a prediction of a first-order
phase transition.%
\footnote
    {%
    For example, as an alternative to the $d = 4{-}\epsilon$ expansion,
    some work has been done using a $d=2{+}\epsilon$ expansion \cite {Hikami}.
    These studies find second-order phase transitions,
    although they are restricted to considering
    theories with $\lambda_1=\infty$.
    }
A useful setting for discussing this point is provided by
the generalization of the single scalar U(1)-Higgs theory to
a theory of $N$ charged scalars with a global U($N$) symmetry.
For $N$ very large, one finds that the flow looks
like Fig.~\ref {fige} instead of Fig.~\ref {figd}.
An infrared-stable fixed point and a tricritical point have appeared.
Theories which start to the
right of the dashed line have second-order phase transitions; those
to the left undergo first-order transitions.
Near four dimensions, the transition between Fig.~\ref {figd}
and Fig.~\ref {fige} occurs at $\Nc(4) = 183.0$.
As the dimension $d$ is decreased and second-order transitions
become more likely, the critical number of fields, $\Nc(d)$, should decrease.
We would like to know whether $\Nc(3) < 1$.

    In section IV, we compute the first sub-leading correction to
$\Nc$ within the $\epsilon$-expansion.
This is only a computation of the slope of $\Nc(d)$ near $d=4$.
However, for both U(1) and SU(2), when extrapolated to $\epsilon=1$,
we find that the $O(\epsilon)$ term is larger than the leading order term
and of opposite sign, so that $\Nc$ naively extrapolates to $\Nc(3) < 1$.
But one cannot trust this result
because the series is clearly badly behaved.
This in itself is somewhat discouraging because $\Nc$
is revealed to be an example of a physical quantity for which the
$\epsilon$-expansion is not very well behaved.

    We do not believe that there is any compelling evidence one way or
the other as to whether the SU(2) phase transition, with one Higgs doublet,
is first or second order for large Higgs mass.
This is an issue to be resolved by lattice simulations or other techniques.
If it is second order for large or moderate $\lambda_1/g_1^2$, then the
$\epsilon$-expansion is doomed.
Our philosophy will be to assume that the transition is indeed
first-order and see what we can extract from the $\epsilon$-expansion.

    Before closing this section, we note one other hurdle for the
$\epsilon$-expansion posed by recent literature.
One can solve the U(1) and SU(2) theories directly in three dimensions
in the limit $N \to \infty$.
Alternatively, one can compute to some
finite order in the $\epsilon$-expansion, set $\epsilon = 1$,
and only then take the $N \to \infty$ limit of the result.
A comparison of the two results gives a test of the $\epsilon$-expansion.
In particular, one can compute the value of $\lambda/g^2$
for the line that divides first- from second-order transitions.
This has been investigated in the large-$N$ expansion by
Jain \etal \cite {Jain}.
Their result is completely different from that
suggested by the $\epsilon$-expansion,
differing even in the power of $N$.
In section IV, we discuss the flaw in their analysis
and redo their calculation.
We find that the critical value of $\lambda/g^2$ scales like
$1/N$ as $N\to\infty$ and that this behavior is in agreement
with the $\eps$-expansion.

\subsection {The magnetic mass and non-perturbative physics}

\label {magnetic mass section}

It is well known that unbroken non-Abelian gauge theories
at high temperature
(such as SU(2) electroweak theory in the symmetric $\phi=0$ phase)
suffer a breakdown of perturbation theory beyond the first few orders.
The loop expansion parameter is not just large; it is infrared divergent.
This reflects the fundamentally non-perturbative nature of long distance
physics in these theories.
In this section, we discuss how this affects the $\eps$-expansion.

\begin {figure}
    {%
    \begin {center}
	\leavevmode
	
	\epsfbox [150 220 500 550] {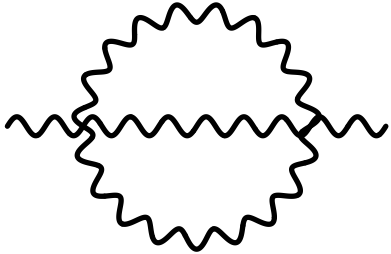}
    \end {center}
    \caption
	{%
	\label {figmag}
        An example of an
        infrared divergent contribution to the transverse gauge field mass
        in three dimensions.
	}%
    }%
\end {figure}

An easy way to see the problem is to
study the effective three-dimensional theory and attempt to
compute the transverse gauge field mass in perturbation theory.
Gauge invariance and rotational invariance imply that
the self-energy of the gauge field $\vec A$ should have the form
\begin {equation}
   \Pi_{ij}(\vec p \,)= (\vec p^{\>2} \delta_{ij} - p_i p_j) f(\vec p^{\>2})
\,.
\end {equation}
Unless $f(\vec p^{\>2})$ singular as $\vec p^{\>2} \rightarrow 0$,
there will be no mass at any order in perturbation theory.
However, fig.~\ref{figmag}\ shows a self energy graph which is
infrared divergent.
This divergence can be seen using simple three-dimensional power counting
which shows this diagram gives the logarithmically-divergent contribution
\begin {equation}
   g_1^4 T^2 \int {\d^6 k\over k^6} ,
   \label {eqpwra}
\end {equation}
at zero external momentum.
Higher order diagrams have increasingly severe divergences.
Non-perturbative fluctuations produce a finite correlation length which
serves to cut-off these divergences.
The inverse correlation length (or magnetic mass) is of order $g_1^2 T$
(or $g^2 \mu$).
If one now computes a {\it three}-loop diagram
using this scale as an infrared cut-off, one finds a contribution of order
\begin {equation}
   g_1^6 T^4 \int\nolimits_{g_1^2 T} {\d^9 p\over p^{10}}
   \sim
   g_1^4 T^2 .
\end {equation}
This is the same size as the two-loop diagram, as
are all higher-loop contributions!
Perturbation theory is therefore
inadequate to treat infrared physics at scales of order $g_1^2 T$.

What happens to this problem in $4{-}\eps$ dimensions?  The two-loop
diagram of fig.~\ref{figmag}\ is no longer infrared divergent when
$\eps < 1$, and infrared problems do not arise until higher order.
In particular, an $n$-loop contribution has the form
\begin {equation}
   (g^2 \mu^\eps)^n \int {\d^{4-\eps}p\over p^{4n-2}} \,,
\end {equation}
so that infrared divergences first appear at order $n=2/\eps$.
Perturbation theory is useless beyond this order, which means
that no quantity can be computed perturbatively with relative
error smaller than $g^n = g^{2/\eps}$.  Since $g^2$ is $O(\eps)$
in our procedure for the $\eps$-expansion, this relative error
is $[O(\eps)]^{1/\eps}$.  But this is just the same order as the
uncertainty intrinsic to the $\eps$-expansion in {\it any\/} theory
due to the fact that the expansion in $\eps$ is an asymptotic
series; terms of order $\eps^k$ in the series start
getting larger beyond order $k \sim 1/\eps$.
So the non-perturbative physics related to the mass gap
in three-dimensional non-Abelian gauge theories
may not pose any problem
for the $\eps$-expansion not already present from
the asymptotic nature of the expansion.

Some physical quantities we will examine are infrared-safe as they depend
only on physics in the asymmetric phase
(where all particles have perturbative masses),
or do not probe the symmetric phase over long distances.%
\footnote
    {%
    The sphaleron mass (or baryon violation rate) in the asymmetric phase
    is an example of the latter case.
    }
Other quantities, such as the scalar correlation length in the symmetric
phase are, in three dimensions,
infrared safe at lowest order but not at higher orders.
How well the $\eps$-expansion will work,
after extrapolating $\eps \to 1$, for such quantities remains unclear.
This issue will be discussed further in section III C.
We will, however, largely focus on infrared-safe quantities
when computing beyond leading order.

\section {One-loop analysis}

\subsection {Renormalization group flow}

We are now ready to look in detail at the solutions to the
one-loop renormalization-group equations.
We shall quote results for both U(1) gauge theory with $N$ charged
U($N$)-symmetric scalars, and SU(2) gauge theory with $N/2$
U($N/2$)-symmetric scalar doublets.%
\footnote
    {%
    More specifically, in the SU(2) theory we require
    Sp$(N)_{\rm R}$ custodial
    symmetry, which is the U($N/2$) flavor-symmetric generalization of
    the usual ${\rm SU}(2)_{\rm R} \simeq {\rm Sp}(2)_{\rm R}$ custodial
    symmetry of the one-doublet case.
    The complete global symmetry
    ${\rm SU}(2)_{\rm L} \times {\rm Sp}(N)_{\rm R}$ implies that the
    scalar potential can only be a function of a single variable,
    $\sum_\alpha |\vec\Phi_\alpha|^2$,
    so that the potential is actually an O($2N$) invariant function of $\Phi$.
    }
$N$ will always denote the number of complex scalar degrees of freedom.
The Lagrangians are%
\footnote
    {%
    We use an unconventional normalization in the Lagrangians
    (\ref {lagrangian});
    a more conventional choice is obtained by rescaling
    the field $\vec\Phi$ by $\sqrt 2$.
    }
\begin {mathletters}
\begin {eqnarray}
      \LE^{\rm U(1)} &=& {\textstyle {1 \over 2}}
                      \left|(\partial - i \mu^{\eps/2} g A) \vec\Phi\right|^2
                    + {\textstyle {1\over 2}} \, m^2 |\vec\Phi|^2
                    + {\textstyle {1\over4!}} \,
                                         \mu^\epsilon\lambda|\vec\Phi|^4
                    + {\textstyle {1\over4}}  \,
			F_{\mu\nu}^2 \,,
\label {langu}
\\
      \LE^{\rm SU(2)} &=& {\textstyle {1 \over 2}}
                        \left|(\partial - i \mu^{\eps/2} g A^a \tau_a /2)
                                         \vec\Phi\right|^2
                    + {\textstyle {1\over 2}} \, m^2 |\vec\Phi|^2
                    + {\textstyle {1\over4!}} \,
                                         \mu^\epsilon\lambda|\vec\Phi|^4
                    + {\textstyle {1\over4}}  \,
			\left( F^a_{\mu\nu} \right)^2 .
\label {langsu}
\end {eqnarray}\label {lagrangian}%
\end {mathletters}%
We will express potentials in terms of the scalar field modulus
$\phi = |\vec\Phi| \equiv (\sum_a \Phi_a^2)^{1/2}$,
in terms of which the classical potential takes the canonical form,
\begin {equation}
\label {canpot}
   V(\phi) = {\textstyle {1\over2}} \, m^2 \phi^2
           + {\textstyle {1\over4!}} \, \mu^\epsilon \lambda \, \phi^4 \,.
\end {equation}
We choose couplings $\lambda$ and $g^2$ to be dimensionless in any
spacetime dimension
and therefore need to include explicit powers of the renormalization
point $\mu$ in the interaction terms.

The renormalization group will be used to run from a renormalization point
$\mu_1 = T$ to a final scale $\mu$.
Our notation will later be
more compact if we write $\mu = \mu_1/s$, where $s$ starts at 1
and grows to $\infty$ as one flows into the infrared.
In place of the gauge couplings in (\ref {lagrangian})
with standard normalizations,
it will also be convenient to use the ``charge'',
\begin {equation}
\label {qdef}
  q \equiv \cases {  g,   & {\rm U(1)} \cr
                    g/2,  & {\rm SU(2)} .  }
\end {equation}
The mass of the gauge boson is then $q \mu^\epsilon \phi$ in both theories.
The introduction of the charge $q$ will allow us to write common
expressions for most one-loop results in either gauge theory.
The one-loop RG equations take the form
\begin {mathletters}
\begin {eqnarray}
        s {d q^2\over d s}
    &=&
        \epsilon \, q^2 - \beta_0 \, q^4 \,,
\label {srngA}
\\
        s {d\lambda\over d s}
    &=&
        \epsilon \, \lambda - (a \, q^4 + b \, q^2 \lambda + c \, \lambda^2)
\,,
\end {eqnarray}\label {srng}%
\end {mathletters}%
where the constants are given by
\begin {eqnarray}
\label {acoeff}
   \beta_0 = {2(N -44 \, C_2) \over 3 (4\pi)^2} \,,
\qquad
   a = {36 \, \ng \over (4\pi)^2} \,,
\qquad
   b = - {12 \, \ng \over (4\pi)^2} \,,
\qquad
   c = {2(N+4) \over 3(4\pi)^2} \,,
\label {betacoeff}
\end {eqnarray}
where
\begin {equation}
   \ng \equiv \cases{ 1 , & {\rm U(1)} ; \cr
                 3 , & {\rm SU(2)} ,  }  \qquad
\label {ngdef}
\end {equation}
is the number of gauge bosons.
$C_2$ is the quadratic Casimir which, for our two gauge groups of interest,
equals $\ng - 1$.
We shall later also need the renormalization group equation
for the mass $m^2$,
\begin {equation}
   \label {mrng}
   s{d m^2\over ds} = - (l_0 \, q^2 + l_1 \, \lambda) \, m^2 \,,
\end {equation}
where
\begin {equation}
   l_0 = - {6 \, \ng\over(4\pi)^2} \,,
   \qquad
   l_1 = {2(N+1)\over 3(4\pi)^2} \,.
\end {equation}
We shall not need the renormalization group equation
for the field $\phi$.

The solution of the RG equation for $q^2$ is
\begin {equation}
   q^2(s) = s^\epsilon q_1^2 \Bigm/
              \Bigl[
                  1 + {\beta_0\over\epsilon} (s^\epsilon{-}1) \, q_1^2
              \Bigr]\,.
\label {eqg}
\end {equation}
The solution for $\lambda$ is most easily found by changing variables
to the ratio $x \equiv \lambda/q^2$ whose RG equation,
\begin {equation}
   s {d x\over d s} = - q^2 [a + (b-\beta_0)x + c x^2] ,
\label {xrng}
\end {equation}
has the solution%
\footnote
    {%
    This solution was previously derived
    for the U(1) case in \cite {Chen&Lubensky&Nelson}.
    }
\begin {eqnarray}
        x(s)
    &=&
        {1\over2c} \, (\beta_0 - b + \sqrt{\Delta} \, \tan\theta(s)) \,,
\label {eqx}
\\
\noalign {\hbox {where}}
        \theta(s)
    &\equiv&
        \alpha
         - {\sqrt{\Delta}\over2\beta_0}
           \ln\left[
               1 + {\beta_0\over\epsilon} (s^\epsilon{-}1) \, q_1^2
             \right] \,,
\label {eqtheta}
\\
        \alpha
    &\equiv&
        {\rm Tan}^{-1}
        \left(
                2cx_1 - (\beta_0-b)\over\sqrt{\Delta}
        \right) \,,
\label {eqalpha}
\\
\noalign {\hbox {and}}
    \Delta &\equiv& 4 a c - (\beta_0-b)^2 \,.
\label {eqDelta}
\end {eqnarray}
This is the form of the solution most convenient for the cases of
Fig.~\ref {figc} or Fig.~\ref {figd},
which occur when $\Delta > 0$ and no stable fixed point exists.
When $\Delta < 0$, two new fixed points appear as shown in Fig.~\ref {fige},
and Eq.~(\ref {eqx}) may be put in the equivalent form:
\begin {eqnarray}
   x(s) &=& {1\over2c} \left[ \beta_0 - b
            + \sqrt{-\Delta} \left(w(s)-1\over w(s)+1\right)
            \right],
\label {eqxb}
\\
\noalign {\hbox {where}}
   w(s) &\equiv& \left(x_1-x_-\over x_+-x_1\right)
              \left[1 + {\beta_0\over\epsilon} \, (s^\epsilon{-}1) \, q_1^2
                           \right]^{\sqrt{-\Delta}/\beta_0},
\label {eqw}
\\
\noalign {\hbox {and}}
   x_{\pm} &\equiv& {1\over2c} \, (\beta_0 - b \pm \sqrt{-\Delta}) .
\label {eqxpm}
\end {eqnarray}
The lines $x=x_+$ and $x=x_-$ pass respectively through the
stable fixed point and the tricritical point shown in Fig.~\ref {fige}.

Return now to the solution (\ref {eqx}) describing
the purely first-order case with which we are mostly concerned.
We want to flow to the region $\lambda \ll q^2$,
where we will use perturbation theory.
For the sake of simplifying formulas and derivations, it is
particularly convenient to run precisely to $\lambda = 0$.
When $\lambda(s)=0$,
the values of the scale factor $s$ and the charge $q^2$
extracted from Eq.~(\ref {eqx}) are%
\footnote
    {%
    In Ref.~\cite {Ginsparg}, it was incorrectly
    asserted that the strength of the phase transition
    is exponentially small when $q_1 \ll 1$ and $\lambda_1/q_1^2 \gsim 1$.
    The assertion was based on examining the form of
    (\ref {xrng}) and noting that for $x \sim 1$ the right-hand side is
    of order $q^2$.
    To run $x$ to 0 requires a change in $x$ of order 1
    and hence, roughly, a change in $\ln s$ of order $1/q^2$.
    Ref.~\cite {Ginsparg} then asserted that $\ln s$ must be order $1/q_1^2$,
    implying that $s$ must run exponentially far before $x$ becomes small.
    This is incorrect because $q(s)$ is not a constant.
    From Eq.~(\ref {eqg}), $q(s)$ runs to be order 1
    when $s^\epsilon$ is order $1/q_1^2$,
    and at that point $x$ can quickly run to zero.
    But $s = O(q_1^{-2/\epsilon})$
    is not exponentially large when $\epsilon\to 1$.
    }
\begin {eqnarray}
   q^2 &=& q_0^2 - (q_0^2-q_1^2) \exp \left\{
                      - {2\beta_0\over\sqrt{\Delta}} \left[
                           {\pi\over2} -
                           {\rm Tan}^{-1} \left(
                              2a-x_1 \, (\beta_0-b)
                              \over x_1 \, \sqrt{\Delta} \right) \right]
                  \right\}
\label {eqgg}
\\
\noalign {\hbox {and}}
   s^{\epsilon} &=& 1 + {q_0^2\over q_1^2} \left( \exp \left\{
                        {2\beta_0\over\sqrt{\Delta}} \left[
                           {\pi\over2} -
                           {\rm Tan}^{-1} \left(
                             2a-x_1 \, (\beta_0-b)
                             \over x_1 \, \sqrt{\Delta} \right) \right]
                   \right\}
                   - 1 \right) ,
\label {eqgs}
\end {eqnarray}
where we have defined
\begin {equation}
\label {eqggo}
   q_0^2 \equiv {\epsilon\over\beta_0} .
\end {equation}
The $x_1 \to \infty$ limits of these formulae were used to derive
the estimate of the scaling factor appearing in our discussion
of the lattice simulations of Dasgupta and Halperin in section
\ref {dasguptasec}.
By taking both the $x_1 \to \infty$ and $q_1 \to 0$ limits,
one can extract the maximum value of $q$ in the shaded region of
Fig.~\ref {figd},
\begin {equation}
   q_{\rm max}^2 = q_0^2 \left( 1 - \exp\left\{
                      - {2\beta_0\over\sqrt{\Delta}} \left[
                           {\pi\over2} +
                           {\rm Tan}^{-1} \left(
                              \beta_0-b\over\sqrt{\Delta} \right) \right]
                   \right\} \right) .
\label {eqggmax}
\end {equation}

\subsection {The one-loop potential}

Having run to $\lambda \ll q^2$, and in particular $\lambda = 0$,
we now want to use ordinary perturbation theory to compute various
quantities describing the phase transition.
The tree-level Higgs potential is
\begin {equation}
\label {vtree}
   V^{(0)}(\phi) = {\textstyle {1\over2}}
                   \left[1 +
                      {1\over\epsilon}(l_0 q^2 + l_1 \lambda - 2\gamma_\phi)
                   \right] m^2 \phi^2
                 + {\textstyle {1\over4!}} \mu^{\epsilon}
                  \left[\lambda +
                      {1\over\epsilon}(a q^4 + b q^2\lambda + c\lambda^2
						    - 4 \lambda\gamma_\phi)
                  \right] \phi^4 ,
\end {equation}
where we have included the one-loop counter-terms for the Minimal
Subtraction (MS) scheme and $\gamma_\phi = \gamma_\phi(\lambda,q^2)$
is the one-loop anomalous dimension of $\phi$.
Working in Landau gauge, the one-loop potential is
\begin {equation}
\label{vvone}
     V^{(1)}(\phi) = V^{(0)}(\phi)
               + {\cal I}(m^2 {+} {\textstyle {1\over2}}\lambda\bar\phi^2)
               + (2N{-}1) \,
                 {\cal I}(m^2 {+} {\textstyle {1\over6}}\lambda\bar\phi^2)
               + \ng \, (3{-}\epsilon) {\cal I}(q^2 \bar\phi^2) \,,
\end {equation}
where it is notationally convenient to introduce the dimension one field
\begin {equation}
\label {phibdef}
   \bar\phi \equiv \mu^{\epsilon/2} \phi .
\end {equation}
The basic one-loop contribution ${\cal I}$ is given by
the (dimensionally regularized) integral
\begin {equation}
        {\cal I}(z)
    =
        {\textstyle {1 \over 2}}
        \int {d^d k \over (2\pi)^d} \> \ln (k^2 + z)
    \equiv
        - {\Gamma (-2 + \epsilon/2 )
                \over 2 \, (4\pi)^{2-\epsilon/2}} \, z^{2-\epsilon/2} .
\label {eqI}
\end {equation}
The arguments of the three ${\cal I}$ functions appearing in Eq.~(\ref {vvone})
are the masses of, respectively, the Higgs boson,
the unphysical Goldstone boson,
and the vector boson in the background of $\phi$.

There are a few technical points to mention before proceeding.
First, we have ignored any $\phi$-independent constants in the effective
potential.
Such constants are not relevant to any of the quantities
we shall be considering; including the constant term would require
that we study its renormalization-group flow as well.
Second, we emphasize again that our convention is $d=4{-}\epsilon$ and not
$d=4-2\epsilon$.
Finally, we have
used MS regularization rather than, say, \MSbar\ renormalization.
Including finite counter-terms to
implement \MSbar\ regularization would not, of course, change the relationship
between physical quantities.
For example, an expansion of critical exponents
of a second-order phase transition in terms of $\epsilon = 4-d$, which
is ``physical,'' would be independent of whether intermediate calculations
were done in the MS or \MSbar\ scheme.
However, in our application final results will also depend on
$\lambda_1$ and $q_1^2$, whose definition are scheme dependent.
But the difference between the MS and \MSbar\ definitions of
$\lambda_1$ and $q_1^2$ is order $O(\lambda_1^2,q_1^4)$,
and hence can be ignored under our working assumption
(\ref {coupling-inequalities})
that $\lambda_1$ and $q_1^2$ are small (with $\lambda_1 \gg q_1^4$).

    Now substitute $\lambda=0$ into the above potential.
Ignoring $\phi$-independent constants, only the vector contribution to the
one-loop potential survives and one finds
\begin {eqnarray}
   \mu^\epsilon V^{(1)}(\phi) &=&
               {\textstyle {1\over2}} \, m^2\bar\phi^2 +
               {\textstyle {1\over 4!}} \, a (q \bar\phi)^4 \, {1\over\epsilon}
                    \left[ 1 - h(\epsilon)
                          \left(q\bar\phi\over\bar\mu\right)^{-\epsilon}
                    \right],
\label {vone}
\\
\noalign {\hbox {where}}
       h(\epsilon)
   &\equiv&
       \epsilon \Bigl(1 {-} {\epsilon \over3} \Bigr)
       \Gamma (-2 + {\textstyle {1\over2}}\epsilon)
           e^{\epsilon \, \gammaE/2},
\label {hdef}
\end {eqnarray}
and where it's convenient to introduce the \MSbar\ scale
\begin {equation}
\label {phimubdef}
   \bar\mu \equiv \sqrt{4\pi\over e^{\gammaE}} \, \mu \,.
\end {equation}
Some useful limits of $h(\epsilon)$ are
\begin {eqnarray}
   h(\epsilon) &=&    1 + {5\over12} \epsilon
                  + \left({3\over16}+{\pi^2\over48}\right)\epsilon^2
                  + O(\epsilon^3) ,
\label {hlimsa}
\\
\noalign {\hbox {and}}
   h(1) &=& {16 \pi \over 9} \sqrt{e^\gammaE \over 4\pi}  \,.
\label {hlimsb}
\end {eqnarray}
Expanded to leading order in $\epsilon$, the one-loop potential is
\begin {equation}
\label {vlead}
   \mu^\epsilon V^{(1)}(\phi) =
               {\textstyle {1\over2}} \, m^2\bar\phi^2 +
               {\textstyle {1\over 4!}} \, a (q \bar\phi)^4
	       \left[
                       \ln\left(q\bar\phi\over\bar\mu\right)
                       - {5\over12}
               \right]
               + O(\epsilon) \,.
\end {equation}
However, for the sake of later tests of the $\epsilon$-expansion, it
will be convenient to keep the potential in the more general form
of Eq.~(\ref {vone}).
Note that substituting $\epsilon=1$ into (\ref {vone}) produces the cubic
$\phi^3$ term familiar from earlier studies of the electroweak phase
transition employing the one-loop effective potential
\cite {baryogenesis,Dine,Carrington}.

    Recall that the three-dimensional mass $m$ is adjusted by changing
the temperature in the original, four-dimensional theory.
At the phase transition, the
potential has two degenerate minima, so that $V(\phic)=V(0)$ and
$(\partial/\partial \phi) V(\phic) = 0$ where $\phic$ is the non-zero
Higgs condensate at the asymmetric minimum.
Applying these conditions to the one-loop potential (\ref {vone}) yields
\begin {eqnarray}
\label {eqvmc}
    \mc^2 &\equiv& {a q^2 \over 4!}
        \left(1-{\epsilon\over2}\right)^{-1} f(\epsilon)^2 \, \bar\mu^2
\\
\noalign {\hbox {and}}
    \bar\phic &\equiv& {1\over q} f(\epsilon) \, \bar\mu \,,
\label {eqphic}
\end {eqnarray}
where
\begin {eqnarray}
    f(\epsilon) &\equiv&
    \left[ \left(1-{\epsilon\over2}\right) h(\epsilon) \right]^{1/\epsilon}
    = e^{-1/12} \left[1 + \left(-{7\over288}+{\pi^2\over48}\right)\epsilon
                 + O(\epsilon^2) \right] .
\label {eqfdef}
\end {eqnarray}
It should be emphasized that these are the values of $m^2(\mu)$
and $\phi(\mu)$ at the scale $\mu$ where $\lambda(\mu) = 0$,
and not at the original scale $\mu_1=T$.

    Before using these results to examine various physical quantities,
we should make a few comments about our choice of scale so that
$\lambda(s) = 0$.
First, recall that the vector contributions to the potential
are the important ones, and so the most important physical
scale which affects the phase transition is the vector mass
$M \sim q \bar\phi$.
One should therefore choose $\mu \sim M$ in order to
avoid the appearance of what in four dimensions would be large
logarithms $\ln(M/\mu)$ in the perturbation expansion and in
$4{-}\epsilon$ dimensions are large factors of
$[(M/\mu)^\epsilon - 1]/\epsilon$.
But, {\it a posteriori}, we see from (\ref {eqphic}) that this is precisely
what we have done.
This is an alternative phrasing for
our previous criteria that perturbation theory
is good when $\lambda/q^2$ is small.
Slightly different choices of $\mu \sim M$
(for example, choosing $\lambda(s) = 0.1 \, q^4(s)$) will give the same results
for physical quantities, order by order in perturbation theory.

    But what would happen if we ran just a tiny bit further until
$\lambda(s) < 0$ while keeping $\lambda/q^2$ small?
The classical potential (\ref {canpot}) is then unbounded below!
This disaster is illusory, however, because both the bare potential
(\ref {vtree}) with one-loop counter-terms and
the one-loop potential (\ref {vvone}) remain bounded below.
It must be
remembered that $\lambda$ in the MS scheme is not itself a physical
quantity, and there is no reason why it can't be negative for
a sensible theory.
This is why the running of $\lambda$ to negative values
was only {\it suggestive} of a first-order transition for small
$\epsilon$.
Only by computing the effective potential, when $|\lambda| \ll q^2$,
can one be sure.

\subsection {Applications}
\label {applications1-sec}

    We are now in a position to put everything together and
compute various quantities of physical interest.
The most interesting qualitative conclusions will follow from
our first two examples, the scalar correlation length and the
rate of baryon number violation.
We will also summarize results for a number of other quantities;
these results will be useful when testing the $\epsilon$-expansion
against unimproved perturbation theory
({\em i.e.}, conventional perturbation theory without the renormalization
group)
at the end of this section.
For the sake of these later tests, it will be helpful to
have on hand full one-loop results in any dimension,
rather than just the first term or two in the expansion around $\epsilon = 0$.

\subsubsection {Scalar correlation length, $\xi$}

As a first example, consider the scalar correlation length
$\xi\sym = 1/\mc$ in the symmetric $\phi = 0$
phase at the critical temperature.
Using the one-loop result (\ref {eqvmc}) for the critical mass,
and the definitions $\mu = \mu_1/s = T/s$, we have
\begin {equation}
\label {eqxisym}
   \xi\sym = {s \over T} \,
           (f(\epsilon) q(s))^{-1}
           \sqrt{{4!\over a}{e^\gammaE\over4\pi}
                           \left(1-{\epsilon\over2}\right)}
           \> \{1 + O(q^2(s))\} ,
\end {equation}
where $q(s)$ and $s$ are given by Eqs.~(\ref {eqgg}) and (\ref {eqgs}).
Similarly, the correlation length
at the critical temperature in the asymmetric phase where $\phi=\phic$
is obtained from the curvature of the one-loop potential at $\phic$:
\begin {equation}
\label {eqxiasym}
   \xi\asym = {s \over T} \,
           (f(\epsilon) q(s))^{-1}
           \sqrt{{4!\over 2a}{e^\gammaE\over4\pi}}
           \> \{1 + O(q^2(s))\} .
\end {equation}

Recall that
our implementation of the $\epsilon$-expansion involves holding
$q_1^2/\epsilon$ and $\lambda_1/\epsilon$ fixed.
Hence $q_1^2$, $q^2(s)$,
$q_0^2$ and $\lambda_1$ are all of order~$\epsilon$.
{}From Eq.~(\ref {eqgs}), it
follows that $s^\epsilon$ is order $\epsilon^0$ and so
the correlation length in either phase depends on $\eps$ as
\begin {equation}
\label {eqxiorder}
  \xi \sim \epsilon^{-1/2} \, e^{O(1/\epsilon)} .
\end {equation}
For small $\epsilon$, the dominant effect is clearly the exponential
dependence of $s$ on $\epsilon$.
The $\epsilon$-dependence is not a simple power series expansion
like we reviewed for critical exponents in section \ref {scalarsec}.
It can be made similar by taking the logarithm:
\begin {eqnarray}
        \ln (\xi\asym T)
    &=&
        {\ln(s^\epsilon) \over\epsilon}
        - {\textstyle {1\over2}} \ln\epsilon
        + \left[
            {\textstyle {1\over2}}
                \ln\left(4! \, \epsilon \over 8\pi a \, q^2(s) \right) +
            {\textstyle {1\over2}} \gammaE + {\textstyle {1\over12}}
          \right]_{\phantom\strut}
        + O(\epsilon)
\label {eqlogxi}
\\
    &=&
        O(\epsilon^{-1}) + O(\ln\epsilon) + O(\epsilon^0) + \cdots.
\nonumber
\end {eqnarray}
Note that terms of $O(\epsilon^0)$ in this formula
which come from the computation of the one-loop potential
are of the same order as the
{\it sub\/}-leading corrections to $s^\epsilon$ in the first term.
So, by the philosophy of the $\epsilon$-expansion, we
should not keep the $O(\epsilon^0)$ terms unless we also compute the
corrections to $s^\epsilon$ using the {\it two\/}-loop renormalization group.
We shall do so in section 3.
Note that the leading $O(1/\epsilon)$ term in (\ref {eqlogxi})
gives nothing more than $\xi \sim s/T$.

    Because this lowest-order calculation does not determine the
overall normalization of the correlation length
({\em e.g.}, it cannot distinguish between
$\xi \sim s / q_0 T$ and $\xi \sim s/(4\pi^2 q_0 T)$),
it is not terribly useful for practical applications.
However, one interesting comparison that is insensitive to an overall
($\epsilon$-independent) multiplicative factor is the ratio of the above
result to what would have been found
with straight one-loop perturbation theory in $4{-}\epsilon$ dimensions
without using the renormalization group.
The exponential dependence on $1/\epsilon$
of the RG unimproved result can be extracted by examining the
one-loop potential (\ref {vvone}) or, equivalently,
by noting that the perturbative
solution to the RG equation (\ref {xrng}) is
\begin {equation}
\label {eqxp}
   x(s) = x_1 - q_1^2 (a + (b-\beta_0)x_1 + c \, x_1^2)
                            \left( s^\epsilon{-}1 \over \epsilon \right)
          + O\!\left[ q_1^4 \left( s^\epsilon{-}1 \over \epsilon \right)
             \right] .
\end {equation}
If only the first two terms in this expansion are retained,
then the solution for $x(s) = 0$ gives
\begin {equation}
    s_{\rm pert}^\epsilon
    =
    1 + {\epsilon \, x_1 \over q_1^2 (a + (b-\beta_0)x_1 + c \, x_1^2)} \,.
\label {eqsp}
\end {equation}

\begin {figure}
\vbox
    {%
    \begin {center}
	\leavevmode
	
	\epsfbox [150 250 500 575] {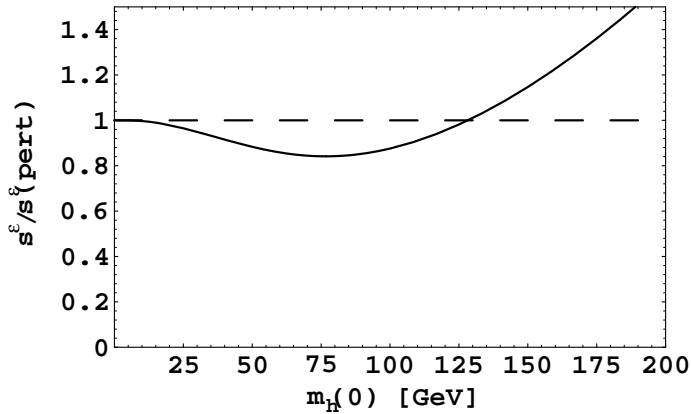}
    \end {center}
    \caption
	{%
	\label {figf}
	The ratio of the RG-improved to unimproved values of the scale factor
	$s^\epsilon$ at one loop order for electroweak theory.
	}%
    }%
\end {figure}

In Fig.~\ref {figf},
we plot the ratio of $s^\epsilon/s_{\rm pert}^\epsilon$ versus
$\lambda_1/q_1^2$
for the SU(2) theory with a single Higgs doublet ($N=2$).
$s^\epsilon$ is the RG-improved result of Eq.~(\ref {eqgs}).
We have fixed $q_1^2/\epsilon$ to be the value $(0.32)^2$
of electroweak theory
and translated $\lambda_1/q_1^2$ into the zero-temperature Higgs mass
$\mh(0)$ using the tree-level relationship
$\mh^2(0) = (4\lambda_1/3g_1^2) \, \mw^2(0)$ with $\mw(0) = 80$~GeV.
Fig.~\ref {figf} suggests that, for light and moderate-mass Higgs bosons
(up to roughly 130 GeV), the phase transition will be
stronger than implied by the unimproved one-loop calculation.
For heavier Higgs bosons, the transition will be weaker.
The increased strength for light mass Higgs is qualitatively
consistent with the results of explicit two-loop calculations in
Refs.~\cite {Arnold&Espinosa,Bagnasco&Dine}.

    Typically, most quantities with positive mass dimension,
such as the latent heat, the Higgs expectation value (in a particular gauge),
or the height of the free energy density barrier between the two vacua,
will scale with a positive power of $s$.
At leading order in the $\epsilon$-expansion,
when compared to unimproved one-loop calculations,
they will therefore have a behavior similar to that shown in Fig.~\ref {figf}.
We shall examine these quantities in more detail later.
First, however, we will examine the potentially more
subtle behavior of the rate of baryon number violation.

\subsubsection {Baryon violation rate, $\Gamma_{\rm B}$}

    The original motivation for this study was to understand whether
baryon number violation is sufficiently slow in the asymmetric phase
to be compatible with electroweak baryogenesis.
For simplicity, we shall study this rate exactly at the critical
temperature where the two vacua are degenerate.
In the case of the real electroweak phase transition,
there is actually a small amount of supercooling below
this temperature before the phase transition completes.

    Since the rate of baryon number non-conservation per unit volume
has positive mass dimension,
it will naively scale with a positive power of $s$
just like the latent heat or the Higgs expectation.
By itself, this suggests that when
the phase transition is stronger than predicted by an
unimproved one-loop calculation,
then the baryon violation rate will also be {\it faster}.
However, this argument is overly simplistic.
The scale factor $s$ is exponentially sensitive to $1/\epsilon$,
which is why the dependence on $s$ dominates the small $\epsilon$
dependence for most quantities.
As we shall discuss, the baryon violation rate has additional
exponential sensitivity to $1/q^2(s)$.
Since $q^2$ is $O(\epsilon)$, this provides an
additional exponential dependence of the rate on $1/\epsilon$
that must be properly accounted for when analyzing the rate to
leading order in the $\epsilon$-expansion.

    The rate of baryon number violation, $\Gamma_{\rm B}$,
is determined by the action of the electroweak sphaleron solution
\cite {Rubakov&Co,Klinkhammer&Manton}.
To compute this rate within the $\epsilon$-expansion,
we need the action of the sphaleron solution in the
$4{-}\epsilon$ dimensional theory.
This poses a problem because the sphaleron is intrinsically a
three-dimensional object, intimately related to topologically
non-trivial mappings of SU$(2) \to S^3$.
There is no obvious generalization of the three-dimensional solution
to a rotationally-symmetric solution in $4{-}\epsilon$ dimensions.
On a concrete level, the problem can be highlighted by considering
the standard form of the sphaleron solution in three dimensions
\cite {Klinkhammer&Manton}:
\begin {equation}
\label {eqspform}
     \vec A(r) = {1\over r} F(r) \, \hat r \times \vec\sigma,
     \qquad
     \Phi(r) = H(r) \, \hat r \cdot \sigma \left(0 \atop 1\right).
\end {equation}
There is no natural generalization of $\hat r \cdot \sigma$ and
$\hat r \times \vec\sigma$ away from three dimensions.

In dimensions larger than three, we may instead consider solutions
which have non-trivial structure in three dimensions and
are translationally invariant in the remaining $d{-}3$ dimensions:%
\footnote
    {%
    This is reminiscent of the contortions one must go through to define
    $\gamma_5$ in dimensional regularization, where four dimensions must
    be treated differently from the others.
    }
\begin {equation}
\label {eqspformb}
     \vec A(\vec x) = \vec A(x_1,x_2,x_3),
     \qquad
     \Phi(\vec x) = \Phi(x_1,x_2,x_3).
\end {equation}
Instead of computing the sphaleron action ${\cal S}_{\rm E}$,
we shall compute the sphaleron action per unit $(d{-}3)$-volume,
${\cal S}_{\rm E}/{\cal L}^{d-3}$,
where we imagine putting the system in a very large box of size~${\cal L}$.

    The action of our sphaleron is determined by the appropriate
stationary point of the functional
\begin {equation}
\label {eqsSa}
   {\cal S}_{\rm E}[\Phi,A] = {\cal L}^{1-\epsilon} \int d^3 x
           \left\{
               {\textstyle {1 \over 2}} \, |D\Phi|^2 +
               {\textstyle {1\over4}} \, F_{\mu\nu}^2 +
               V(\phi)
           \right\} .
\end {equation}
As before, we shall first run to $\lambda(s)=0$ so that perturbative
corrections will be small when $\epsilon$ is small.
Note that at this point the one-loop potential,
as given by Eqs.~(\ref {vone}) and (\ref {eqvmc}),
depends on the coupling $q(s)$ only in the combination $q\phi$.
We can make explicit all the
parameter dependence 
at $\Tc$ by writing
$V^{(1)}(\phi) = \mu^{4{-}\epsilon} v(q \phi / \mu^{1-\epsilon/2}).$
Now rescale all fields and dimensions,
\begin {equation}
\label {rescale}
   A \to {\mu^{1-\eps/2} \over q} A,
   \qquad
   \Phi \to {\mu^{1-\epsilon/2}\over q} \Phi,
   \qquad
   x \to {x \over \mu},
\end {equation}
to produce
\begin {equation}
\label {eqSscaled}
   {\cal S}_{\rm E} = {(\mu {\cal L})^{1-\epsilon}\over q^2} \int d^3 x
   \left\{
          {\textstyle {1 \over 2}}
              \left|
                  \Bigl(\nabla_i - i {\vec \tau} \cdot \vec A_i \Bigr) \Phi
              \right|^2
          + {\textstyle {1\over4}}
              \Bigl(\nabla_i \vec A_j - \nabla_j \vec A_i
                  + \vec A_i \times \vec A_j \Bigr)^2
          + q^2 v(\phi)
     \right\} .
\end {equation}
Recall that $q^2(s)$ is $O(\epsilon)$.
{}From this form of the action it is apparent that
the details of the one-loop potential $v(\phi)$ do not affect
the sphaleron action at leading order in $\epsilon$.
However, the solution is already known in the case where
the strength of the potential is negligible: it is the standard
sphaleron solution for the classical potential $\lambda(\phi^2 {-} v^2)^2$
when $\lambda \to 0$.
The action of the three dimensional sphaleron solution in this limit is
${\cal S}_{\rm E} = 3.0405 \, \mw / \alpha_{\rm w}$
\cite {Klinkhammer&Manton,Yaffe}.
In our case, this translates to
\begin {eqnarray}
        {\cal S}_{\rm E}
    &=&
        3.0405 \, (\mu {\cal L})^{1-\epsilon} \> {\pi \bar\phic \over q \mu}
        \> (1 + O(\epsilon))
\nonumber
\\
    &=&
        3.0405 \, (\mu {\cal L})^{1-\epsilon} \> {\pi e^{-1/12} \over q^2}
        \sqrt{4\pi \over e^\gammaE}
        \> (1 + O(\epsilon))
\nonumber
\\
    &=&
        {23.344 \over q^2} \, (\mu {\cal L})^{1-\epsilon}
        \> (1 + O(\epsilon)).
\label {eqsph}
\end {eqnarray}
The only importance of the potential is to determine the asymptotic
Higgs expectation value $\bar\phic$ through Eq.~(\ref {eqphic}).

\begin {figure}[bh]
\vbox
    {%
    \begin {center}
	\leavevmode
	
	\epsfbox [150 250 500 540] {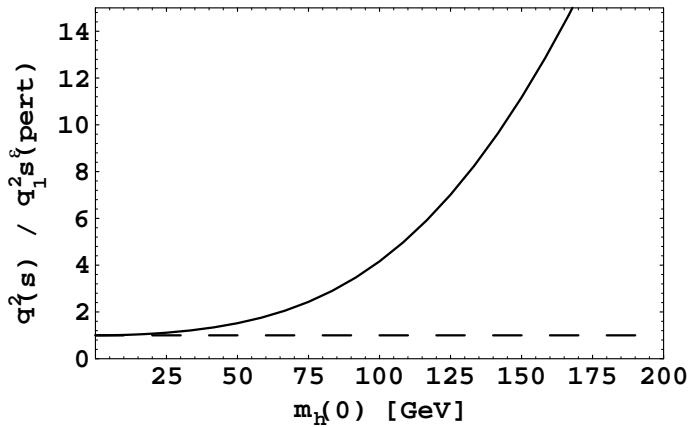}
    \end {center}
    \caption
	{%
	\label {figg}
	The ratio of the RG-improved to the unimproved value of
	the gauge coupling $q(s)$ at one loop order for electroweak theory.
	}%
    }%
\end {figure}

\begin {figure}
\vbox
    {%
    \begin {center}
	\leavevmode
	
	\epsfbox [150 250 500 500] {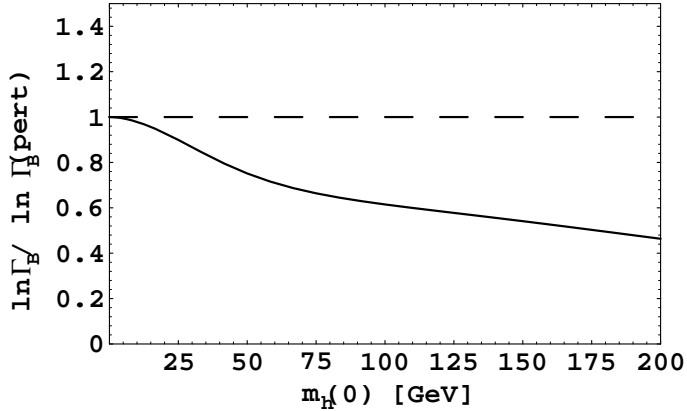}
    \end {center}
    \caption
	{%
	\label {figh}
	The ratio of the RG-improved to the unimproved value of
	the logarithm of the baryon violation rate at one loop order.
	}%
    }%
\end {figure}

    In three spatial dimensions, the rate of baryon number violation
per unit correlation time, $\xi\Gamma_{\rm B}$,
is roughly $({\cal V}/ \xi^3) \exp(-{\cal S}_{\rm E})$ where ${\cal V}$
is the volume of space and $\xi$ is the correlation length.
(The distinction between the scalar and vector correlation lengths,
and the presence of additional prefactors to the exponential,
will not matter at leading order in $\epsilon$.)
For the moment, let us blithely ignore the
$(\mu {\cal L})^{1-\epsilon}$ factor in Eq.~(\ref {eqsph}) and
recall that $\xi \sim s/T$.
Then the leading-order result in the $\epsilon$-expansion is
\begin {equation}
\label {rate}
   {\Gamma_{\rm B} \over \cal V} = T^4 \exp\left[
                - {23.344 \over q^2(s)}
                - {3\over\epsilon} \ln(s^\epsilon)
                + O(\epsilon^0)
             \right] ,
\end {equation}
with $q^2(s)$ and $s^\epsilon$ given by Eqs.~(\ref {eqgg}) and (\ref {eqgs}).
Consider comparing this result to a straight perturbative calculation
that was not improved by the renormalization group.
We have already discussed this comparison for the explicit dependence on $s$.
For $q^2(s)$, the corresponding unimproved result is
$s_{\rm pert}^\epsilon q_1^2$.
A plot of $q^2(s)/(s_{\rm pert}^\epsilon q_1^2)$ versus
$\lambda_1/q_1^2$ is given in Fig.~\ref {figg}.
Inclusion of the renormalization group flow always increases $q^2(s)$,
which drives the rate $\Gamma_{\rm B}$ larger.
In Fig.~\ref {figh}, we put everything together and plot the ratio of the
exponent in Eq.~(\ref {rate}) to the result from
unimproved perturbation theory.


Our leading-order result of Fig.~\ref {figh} suggests
that the baryon violation rate is always larger
than that deduced from one-loop perturbation theory.
If this is a correct qualitative description of what happens
when $\epsilon \to 1$, then the one-loop bound of $\mh(0) < 35$~GeV
derived by Dine \etal \cite {Dine}, is indeed
an upper bound on the Higgs mass for
electroweak baryogenesis in the minimal standard model.

Having reached this punch line, we must return to an annoying technical
point concerning the $(\mu {\cal L})^{1-\epsilon}$ factor in Eq.~(\ref {eqsph})
which we ignored.
Remember that $\mu = T/s$,
so this factor is exponentially sensitive to $1/\epsilon$
but equals unity at $\epsilon = 1$.
If we retain the factor, it
changes the relative importance of the ${\cal S}_{\rm E}$ term and the $\ln s$
term in the exponent of (\ref {rate}) in the $\epsilon$-expansion.
Also, for $\eps<1$, these factors have different dependencies on the
size ${\cal L}$ of the system in the extra $1{-}\eps$ dimensions,
and so are not comparable for ${\cal L} \to \infty$.
When one works to higher orders in the $\epsilon$-expansion
some corrections to $\ln \Gamma_{\rm B}/{\cal V}$ will be proportional
to $(\mu {\cal L})^{1-\eps}$, and serve to modify the sphaleron action,
while other corrections are independent of $\cal L$ and modify
the zero-mode prefactor.
This suggests a natural approach for defining the $\epsilon$-expansion
of the full baryon violation rate:
separately extrapolate to $\epsilon=1$ the terms independent of $\cal L$,
and the (coefficient of) terms proportional to $(\mu {\cal L})^{1-\eps}$.
Combine the two contributions to $\ln \Gamma_{\rm B}$ only at $\epsilon = 1$
where $(\mu {\cal L})^{1-\eps}$ may be replaced by unity.
This procedure, to lowest order, yields Eq.~(\ref {rate})\
(evaluated at $\epsilon=1$) and the results shown in Fig.~\ref {figh}.%
\footnote
    {%
    The qualitative behavior of the ratio plotted in Fig.~\ref {figh}
    is, in fact, insensitive to the presence or absence of the
    $3 \ln s$ zero-mode contribution.
    }

    Performing a complete next-to-leading order calculation
remains a problem for the future.
However, as an example of corrections to the sphaleron action,
consider the one-loop correction due to quantum fluctuations in
modes other than the translational zero-modes.
The prefactor to $\exp (-\SE)$ depends on the determinant of the curvature
operator $\Omega^2_{\rm sph}$
describing small fluctuations about the sphaleron solution.
The non-zero mode contribution to the prefactor has the form%
\footnote{%
   For a discussion of these determinants in the three dimensional case,
   see, for example, Refs.~\cite{Arnold&McLerran} and \cite{Carson} and
   references therein.
}
\begin {equation}
  {\cal P} = \left(
		{\rm Det}'_{4-\eps}(\Omega^2_{\rm sph})
		\over
		{\rm Det}_{4-\eps}(\Omega^2_{\rm vac})
	    \right)^{-1/2}
  = \exp \left\{
        - {\textstyle {1\over2}} L^{1-\eps} \int d^{1-\eps}\omega {\sum_i}
	\ln
	\left[
	    { \omega^2 + \lambda_{\rm sph}^{(i)} \over
	      \omega^2 + \lambda_{\rm vac}^{(i)} }
        \right] \right\} \,,
\end {equation}
where $\Omega^2_{\rm vac}$ is the corresponding curvature
operator around the ground state, the determinants are taken over
fluctuations in $4{-}\eps$ dimensions, $\vec\omega$ is the momentum
of a fluctuation in the $1{-}\eps$ transverse dimensions in which the sphaleron
is static, and $\{ \lambda^{(i)} \}$ are the eigenvalues of $\Omega^2$
in {\it three} dimensions.
The prime on the determinant indicates
the omission of the zero eigenvalues, which can't be treated
by Gaussian integration.  Since the only scale in the problem is $\mu$,
this prefactor must have the form of
\begin {equation}
   {\cal P} \sim \exp\{ (\mu {\cal L})^{1-\eps} f(q,\eps) \}.
\end {equation}
The exponent has the same $(\mu {\cal L})^{1-\eps}$
factor as does the sphaleron action in (\ref{eqsph}), and so $\ln {\cal P}$
generates an $O(q^2) = O(\eps)$ correction to the classical action.

Another possible approach for defining the $\epsilon$-expansion of the
baryon violation rate is to
choose the length ${\cal L}$ of the extra $1 {-} \epsilon$ dimensions
to be order $\mu^{-1}$.
This allows the zero-mode prefactors to
be treated on the same footing as the action and other prefactors,
but at the price of introducing an unphysical dimensionless parameter,
$\mu {\cal L}$.%
\footnote
    {%
    This approach also has an advantage of allowing one to generalize
    the sphaleron determinant to $4{-}\eps$ dimensions
    while retaining only a single negative mode.
    In contrast, taking ${\cal L} \to \infty$ generates a continuum of
    negative eigenmodes which contribute an imaginary part to $\ln \cal P$
    proportional to $i (\mu {\cal L})^{1-\eps} \pi/2$.
    This produces the expected prefactor of $i$,
    reflecting a single unstable mode, only when $\epsilon \to 1$.
    }
Any calculation to finite order in $\eps$ would depend on the precise
choice of $\mu {\cal L}$, in direct analogy to the dependence of calculations
on the renormalization scale in conventional perturbation theory.
As in conventional perturbation theory, the result should become less
and less sensitive to the precise choice of $\mu {\cal L}$ when one computes to
higher and higher orders in $\eps$.
Which approach will yield the best systematic approximation for the
baryon violation rate is not yet clear.
We emphasize that the complications discussed here are specific to the
baryon violation rate and will not be present for other quantities discussed
below.

\subsubsection {Latent heat, $\Delta Q$}

The latent heat (per unit volume) of the phase transition is defined as
the difference in energy density of the two coexisting phases at $\Tc$,
\begin {equation}
   \Delta Q \equiv \Delta E \Bigr|_{T=\Tc} = T \Delta S \Bigr|_{T=\Tc} =
   -T {d\Delta F \over d T} \biggr|_{T=\Tc} \,.
\end {equation}
Here, $\Delta E$, $\Delta S$, and $\Delta F$ denote respectively
the energy, entropy, or free energy densities
in the symmetric ($\phi=0$) phase minus that in the
asymmetric ($\phi=\phic$) phase.
The free energy density $F$ is simply $T\Veff$,
where $\Veff$ is the effective potential in the three-dimensional theory.
Understanding the dominant sources
of temperature dependence in $\Veff$ requires returning to the
relationship between the effective three-dimensional theory and the
original four-dimensional theory at high temperature.
There is explicit dependence from the initial choice of renormalization
scale $\mu_1$ in the three-dimensional theory as equal to the temperature $T$.
There is also implicit dependence from the variation of the
effective three-dimensional mass $m_1 = m(\mu_1)$ with $T$,
as was shown explicitly for pure scalar theory in (\ref{mrenorm}).
In gauge theories, this relationship has the form
\begin {equation}
   m_1^2 = - \nu^2 + (c_q \, q^2 + c_\lambda \, \lambda)T^2
         + O(q^3 T^2, \lambda^{3/2} T^2) ,
   \label {eqexb}
\end {equation}
where $c_g$ and $c_\lambda$ are constants.
Other sources of temperature dependence, such as the running
of the original four-dimensional couplings or the coefficients
of higher-dimensional interactions in the three-dimensional theory,
will not contribute at leading order.
Now, on dimensional grounds, the three-dimensional result for
$\Delta\Veff$ must have the form $\mu_1^3 \, f(m_1^2/\mu_1^2)$.
Combining this with Eq.~(\ref{eqexb}), plus the condition that
$\Delta\Veff(\Tc)=0$ (by definition of $\Tc$),
allows one to rewrite the latent heat as
\begin {equation}
   \label {DQrelation}
   \Delta Q = -T^2
   \left(
       {\partial \over \partial \mu_1} +
       {\partial m_1^2\over\partial T} {\partial \over \partial m_1^2}
   \right)
        \Delta\Veff \biggr|_{\mu_1=T=\Tc}
   =
       {\nu^2 T\over m_1^2} \, \mu_1 { \partial \over \partial\mu_1}
         \Delta\Veff \biggr|_{\Tc} \,,
\end {equation}
neglecting corrections of order $g_1$ or $\sqrt \lambda_1$.
The mass $m_1$ should be held fixed in the above derivative.

We can now turn to the computation of the (rescaled) latent heat
$\Delta Q/\nu^2 T$ in $4{-}\eps$ dimensions.
We have previously determined $m(\mu)$ at the phase transition
(Eq.~(\ref{eqvmc})).
The mass $m_1$ is related to $m(\mu)$ by the renormalization group equation
(\ref{mrng}), whose solution is
\begin {eqnarray}
   m^2(\mu) &=& m_1^2 \> P(s^\eps) \,,
\\
\noalign {\hbox {where at one loop}}
\label{Pdef}
   P(s^\eps) &=&
              \exp\left[ - {1\over\eps}
                \int\nolimits_1^{s^\eps} {d {s'}^\eps \over {s'}^\eps} \,
                \Bigl( l_0 \, q^2(s') + l_1 \, \lambda(s') \Bigr)
                \right] \,.
\end {eqnarray}
Unfortunately, we do not have a simple closed form for this integral.
Note that the ratio $P(s^\eps) = m^2(\mu) / m_1^2$ is $O(\eps^0)$.
Using the one-loop potential
(\ref{vone}), one finds
\begin {eqnarray}
       {\Delta Q\over \nu^2}
   &=& {T^{3-\eps} \over s^{2 - \eps} q^2(s)} {4\pi\over e^\gammaE}
        f(\eps)^2 P(s^\eps)
        \> \{1 + O(q^2(s))\}
\nonumber
\\
   &=& 5.972 \, {T^{3-\eps} \over s^{2 - \eps} q^2(s)}
       P(s^\eps)
       \> (1 + O(\eps)) .
\label {moo}
\end {eqnarray}
This is exponentially sensitive to $1/\eps$ because of the dependence
on $s$.
As with the correlation length, the overall numerical factor
in (\ref{moo}) is not useful unless $s^\eps$ is computed using
the 2-loop renormalization group.
Phrased another way, the logarithm is
\begin {equation}
   \ln\left( \Delta Q \over \nu^2 T^{3-\eps} \right)
   = {-2 + \eps \over \eps} \ln(s^\eps)  -  \ln \eps
     + \left[ \ln\left( 4\pi\eps P(s^\eps) \over q^2(s) \right) - \gammaE -
                 {\textstyle {1\over6}}
        \right] + O(\eps) ,
   \label {eqexe}
\end {equation}
and a 2-loop calculation of $s^\eps$ is needed to
evaluate this result consistently through $O(\eps^0)$.
However, the need for the two-loop renormalization group
is avoided if one computes
the latent heat scaled by an appropriate power of the correlation length,
\begin {equation}
   {\xi\asym^{2-\eps} \, \Delta Q \over \nu^2}
   = T \> {4! \, P(s^\eps) \over 2 a \> q^4(s)} \> (1 + O(\eps)) \,.
\end {equation}

\subsubsection {Free energy difference, $\Delta F(T)$}

We now consider the free energy difference between the symmetric and
asymmetric states as the temperature moves away from $\Tc$.
It is convenient to use a dimensionless reduced temperature
which we will define as
\begin {equation}
   t \equiv {T - T_0 \over \Tc - T_0} \,,
\end {equation}
where $T_0$ is the temperature below $\Tc$ at which $\phi = 0$ is no
longer metastable.
In terms of the three-dimensional theory, this corresponds to using
\begin {equation}
   m^2 = t \, \mc^2 \,,
\end {equation}
(neglecting higher order corrections in $g_1^2$
and assuming that $|t| \lsim 1$).
We shall not dwell on the computation of $\Tc - T_0$, which involves
details of the connection between the original four-dimensional and
effective three-dimensional theories that are beyond the scope of our
present study.

To find $\Delta F(t) \equiv V_{\rm eff} (t,0) - V_{\rm eff} (t,\vev)$~~%
(where $\vev$ is the expectation in the asymmetric phase at temperature $t$),
it is convenient to write $\phi$ in terms of
the rescaled field
\begin {equation}
   \omega \equiv f(\eps) \, q \, \bar\phi / \bar \mu .
\label {omdef}
\end {equation}
The one-loop potential then gives
\begin {eqnarray}
   \Delta F(t)
   &=& {T^{5-\eps} \over s^{4-\eps}} \, {a \over 4! \cdot 4}
       \left( 4\pi \over e^\gammaE \right)^2
       \left(1 - {\eps\over4}\right)^{-1+4/\eps}
       h(\eps)^{4/\eps} D(t,\eps)
       \; \{1 + O(q^2(s))\}
\nonumber
\\
   &=& 0.2302 ~ \ng \, T^{5-\eps} s^{\eps-4}
              D(t, \eps\to 0) \; (1+O(\eps)) \,,
\end {eqnarray}
where
\begin {equation}
   D(t, \epsilon)
   \equiv
   -\left(4-2\eps\over 4-\eps\right)^{-1+4/\eps}
   \min_\omega
   \left(
       2 t \, \omega^2
       - {4\over\eps} \, \omega^{4-\eps}
       + {4-2\eps \over \eps} \omega^4
   \right) \,,
\end {equation}
is the value of the right hand side at its non-zero local minimum.
Note that $D(0, \eps) = 1$, $D(1, \eps) = 0$,
and $D(t,\eps)$ is smooth as $\eps \to 0$.

Once again,
the free energy density difference $\Delta F(t)$ is exponentially sensitive
to $\epsilon$,
but the free energy difference per unit correlation volume,
\begin {equation}
    \xi\asym^{4-\epsilon} \Delta  F(t) =
    {\textstyle {3\over2}} \, {e \, T \over a \,  q^4(s)} \> D(t,\epsilon\to0)
        (1 + O(\eps)) \,,
\end {equation}
is $O(\epsilon^{-2})$.

\subsubsection {Surface tension, $\sigma$}

    At the transition temperature $T_c$,
large domains of differing phases can coexist.
Planar or nearly planar boundaries separating such domains will have
a surface tension (or domain wall energy density) given,
to leading order, by
\begin {equation}
   \sigma = T \int\nolimits_0^{\phic} d\phi \> \sqrt{2 V(\phi)}.
\end {equation}
Using the change of variables (\ref{omdef}), this becomes
\begin {eqnarray}
\nonumber
   \sigma
   &=& {T^{4-\eps} \over s^{3-\eps}} \, \left(4\pi\over e^\gammaE\right)^{3/2}
       \left( a / \epsilon \over 4! \, q^2(s)\right)^{1/2}
       \left(1-{\eps\over2}\right)^{-1/2} f(\eps)^3
\\
   && \qquad\qquad \times
       \int\nolimits_0^1 d\omega\, \omega
           \sqrt{\epsilon - 2 \, \omega^{2-\eps} + (2 {-} \eps) \, \omega^2}
       \> \{\, 1 + O(q(s))\}
\\
   &=& 0.2987 \, \sqrt {\ng} \, {T^{4-\eps} \over s^{3-\eps} q(s)}
       ~(\, 1 + O(\eps)) \,.
\end {eqnarray}

\subsubsection {Bubble nucleation rate, $\GammaN$}

  The bubble nucleation rate $\GammaN$ is of order $\mu^4 \exp(-\SE)$
where $\SE$ is the action of the Euclidean bounce solution
\cite {Coleman}.
The rate increases from zero as the temperature drops below $\Tc$.
The relevant part of the action is the scalar contribution,
\begin {equation}
   \SE = \int d^{4-\eps}x
       \left( {\textstyle {1\over2}}(\partial\phi)^2 + V(\phi) \right).
\end {equation}
We shall look for O($4{-}\eps$)-symmetric bounce solutions.
As discussed earlier, we may parameterize the deviation of the temperature
from $\Tc$ by writing $m^2 = t \, m\c^2$.
The one-loop potential at $\lambda(s) = 0$ has the form
$V^{(1)}(\phi) = \mu^{4-\eps} \, v(q\phi/\mu^{1-\eps/2}, t)$.
The rescaling
\begin {equation}
   \phi \to {\mu^{1-\eps/2}\over q}\phi,
   \qquad
   x \to {x\over\mu q} \,,
\end {equation}
moves the dependence on $q$ out in front of the action,
\begin {equation}
   \SE \to {1\over q^{4-\eps}} \int d^{4-\eps}x
       \left(
           {\textstyle {1\over2}} (\partial\phi)^2 + v(\phi,t)
       \right) \,.
\end {equation}
We shall not numerically compute the extremum of the integral above,
which depends on $\eps$ and $t$.
For small $\eps$ and fixed $t$, however,
the $q^{-4+\eps}$ dependence of the action, together
with fig.~\ref{figg}, implies that the nucleation rate is larger than
would have been found with an unimproved one-loop calculation.

Instead of computing the action of an $O(4{-}\eps)$ symmetric bounce,
one could also consider an $O(3)$ symmetric solution in analogy with
the way we previously treated sphalerons.

\subsection {Testing the $\eps$-expansion when $\lambda_1 \ll q_1^2$}

The motivation for exploring the $\eps$-expansion was the breakdown of
perturbation theory in three dimensions.
When $\lambda_1 \ll q_1^2$, however,
the unimproved loop expansion is a reliable method for studying the
phase transition directly in three dimensions.
This provides an opportunity to test the $\eps$-expansion
in a regime where answers are already known by another technique.
In this section, we shall assume that $\lambda_1 \ll q_1^2$
and expand our previous results in powers of $x_1 = \lambda_1/q_1^2$.
To simplify the expansion, we shall also assume in this section
that the initial three-dimensional couplings
$q_1^2$ and $\lambda_1$ are small while $\lambda_1/q_1^2$
is fixed.
This corresponds to the formal limit (\ref{g-assumptions})
discussed in the introduction.
When generalized to $4{-}\eps$ dimensions, our limit becomes
\begin {equation}
   {q_1^2 \over \eps}, \;\; {\lambda_1 \over \eps} \rightarrow 0,
   \quad \hbox {with} \quad
   {\lambda_1 \over q_1^2} \ll 1 ~ \hbox{and fixed}.
\end {equation}
The couplings
$q_1^2$ and $\lambda_1$ are still numbers of $O(\eps)$, but they
are assumed to be {\it small\/} numbers times $\eps$.
In particular, we shall ignore $q_1^4$ compared to $\eps\lambda_1$.

In this limit, our results (\ref{eqgg}) and (\ref{eqgs}) for $q^2(s)$
and $s^\eps$ at $\lambda(s)=0$ become
\begin {equation}
   q^2(s) \sim s^\eps q_1^2 \sim {\eps \lambda_1\over a q_1^2}
   \ll 1 ,
\label {george}
\end {equation}
which can also be extracted directly from the $x\ll 1$ limit of the
renormalization group equations (\ref{srng}) and (\ref{xrng}).
Since $q^2(s)$ is small,
a one-loop calculation will be good to leading order in {\it any} dimension.

    Consider the one-loop result (\ref{eqxiasym}) for the scalar correlation
length in the asymmetric phase, where we had not yet assumed $\eps$ to be
small.
In any dimension, this result will be valid to leading order
in $\lambda_1/q_1^2$.  Expanding in powers of $\eps$ gives
\begin {eqnarray}
   \xi\asym &=& {1\over T} \left(q_1^2 \over \eps\lambda_1\right)^{1/2}
   \left(\eps\lambda_1\over a q_1^4\right)^{1/\eps}
   \bigl[ 1.41748 - 0.25701\,\eps + 0.10441\,\eps^2 - 0.03710\,\eps^3
\nonumber \\ && \qquad\qquad\qquad\qquad\qquad\qquad {}
          + 0.01760\,\eps^4 - 0.00737\,\eps^5 + O(\eps^6)
          + O(x_1) \bigr] \,.
\label {eqexs}
\end {eqnarray}
We have chosen to factor out the leading power of $\epsilon$ and the
exponential dependence on $1/\eps$
and then written the remainder as a series in $\eps$ starting at $\eps^0$.
Alternatively, we could have expanded $\ln(\xi T)$
in powers of $\eps$ as in (\ref{eqlogxi}).
We shall generally present expansions in the form shown above, however,
for little more reason than that we find the formulas more aesthetic.

Consider the error which is introduced by truncating the series in
(\ref {eqexs})
before setting $\epsilon$ to 1.
Truncating at $O(\epsilon^0)$ and
neglecting all terms proportional to positive powers of $\epsilon$
yields an overestimate of $\xi_c$ which is 14\% larger than
the correct three-dimensional value.
If one instead keeps terms through $O(\eps)$,
then the result is 6\% too small.
Clearly, the $\epsilon$-expansion for the correlation length
is quite well-behaved, at least for small $\lambda_1 / g_1^2$.
For arbitrary $\lambda_1/g_1^2$,
keeping terms to $O(\eps)$ requires a two-loop calculation of
$\xi\c$ and the {\it three}-loop renormalization group.
Unfortunately, the three-loop $\beta$-function for $\lambda$ is
not yet known in either SU(2) or U(1) theory.

    Expansions like (\ref{eqexs}) for some of the other quantities
computed earlier are:
\begin {eqnarray}
   \xi\sym &=& {1\over T} \left(q_1^2 \over \eps\lambda_1\right)^{1/2}
   \left(\eps\lambda_1\over a q_1^4\right)^{1/\eps}
   \!\! \Bigl[ 2.00462 - 0.86462\,\eps + 0.17588\,\eps^2 - 0.09368\,\eps^3
\nonumber \\ && \>\;\qquad\qquad\qquad\qquad {}
          {} + 0.03133\,\eps^4 - 0.01698\,\eps^5 + O(\eps^6)
          + O(x_1) \Bigr]
\\
   \Delta Q &=& \ng \nu^2 T^{3-\eps}
   \left(q_1^2 \over \eps\lambda_1\right)
   \left(\eps\lambda_1\over a q_1^4\right)^{-(2-\eps)/\eps}
   \!\! \Bigl[ 1.36153 + 0.49372\,\eps - 0.06630\,\eps^2 - 0.00538\,\eps^3
\nonumber \\ && \>\qquad\qquad\qquad\qquad\qquad\qquad {}
          {} - 0.00509\,\eps^4 - 0.00021\,\eps^5 + O(\eps^6)
          + O(x_1) \Bigr]
\label {delQ0}
\\
   \sigma &=& \ng T^{4-\eps}
   \left(q_1^2 \over \eps\lambda_1\right)^{1/2}
   \left(\eps\lambda_1\over a q_1^4\right)^{-(3-\eps)/\eps}
   \!\! \Bigl[ 0.14260 + 0.09226\,\eps + 0.00692\,\eps^2 - 0.00185\,\eps^3
\nonumber \\ && \;\qquad\qquad\qquad\qquad\qquad\qquad {}
          {} - 0.00069\,\eps^4 - 0.00028\,\eps^5 + O(\eps^6)
          + O(x_1) \Bigr]
\\
   \Delta F(T_0) &=& \ng T^{5-\eps}
   \left(\eps\lambda_1\over a q_1^4\right)^{-(4-\eps)/\eps}
   \!\! \Bigl[ 0.23025 + 0.31089\,\eps + 0.19796\,\eps^2 + 0.10668\,\eps^3
\nonumber \\ && \>\qquad\qquad\qquad\qquad {}
          {} + 0.05882\,\eps^4 + 0.03215\,\eps^5 + O(\eps^6)
          + O(x_1) \Bigr]
\label {delFexp}
\end {eqnarray}
The multiplicative errors made by keeping only terms through $O(\eps^0)$
or $O(\eps^1)$ in the brackets are shown in table~\ref{tableb}.
Note that the result for $\Delta F(T_0)$ is especially sensitive to the
$O(\eps^1)$ term.

\begin{table}
\begin {center}
\tabcolsep=8pt
\begin {tabular}{|lc|cc|}             \hline
\multicolumn{1}{|c}{observable ratio}
  &                   & LO          & NLO         \\ \hline
asymmetric correlation length
  &   $\xi\asym$      & 1.14        & 0.94        \\
symmetric correlation length
  &   $\xi\sym$       & 1.62        & 0.92        \\
latent heat
  &   $\Delta Q$      & 0.77        & 1.04        \\
surface tension
  &   $\sigma$        & 0.60        & 0.98        \\
free energy difference
  &   $\Delta F(T_0)$ & 0.24        & 0.56        \\ \hline
\end {tabular}
\end {center}
\caption
    {%
    \label {tableb}
    The ratio of the $\eps$-expansion results,
    computing prefactors through leading order (LO) and
    next-to-leading order (NLO) in $\eps$,
    to the corresponding three-dimensional result when $\lambda_1 \ll q_1^2$.
    }%
\end{table}

    For most quantities,
we find that a calculation through $O(\eps^0)$
gives agreement to within a factor of 2,
and inclusion of $O(\eps^1)$ terms
yields agreement within 10\%.
There is no guarantee, of course, that the convergence
might not be worse when $\lambda_1/q_1^2$ is large.%
\footnote
    {%
    The series shown in Eqs.~(\ref {eqexs}-\ref {delFexp})
    are actually convergent
    at $\epsilon =  1$ because there are no singularities
    within $|\epsilon| \le 1$ in the relevant formulas such as
    Eq.~(\ref {eqxisym}).
    This is only true at one-loop order; as shown later,
    higher order results do have singularities at $\epsilon = \pm 1$.
    }
The expansion of the free energy difference
is notably worse than the other observables.
In this particular case, the behavior of the series for the
logarithm, $\ln (\Delta F(T_0)/T^{5-\eps})$, is rather different;
next-to-leading order in $\epsilon$ reproduces the correct
three dimensional answer to within 17\%.
We have not shown the expansion of the baryon number violation rate
because, due to the way we constructed its $\epsilon$-expansion,
the result is trivially the same as the three-dimensional answer
when $\lambda_1 \ll q_1^2$.

\begin{table}
\begin {center}
\tabcolsep=8pt
\begin{tabular}{|l|cccccc|}                                       \hline
\multicolumn{1}{|c|}{observable}
 & C$_0$ & RG$_1$ & C$_1$ & RG$_2$ & C$_2$ & RG$_3$            \\ \hline
$\ln\xi\c$, $\ln\Delta Q$, $\ln\Delta F$, $\ln\sigma$
 & --             & $O(\eps^{-1})$ & $O(\eps^0)$
                  & $O(\eps^0)$    & $O(\eps^1)$ & $O(\eps^1)$ \\
$\ln(\xi\c^2\Delta Q)$, $\ln(\xi\c^3\Delta F)$, $\ln(\xi\c^3\sigma)$
 & --             & $O(\eps^0)$    & $O(\eps^0)$
                  & $O(\eps^1)$    & $O(\eps^1)$ &             \\
$\ln\Gamma_{\rm B}$
 & $O(\eps^{-1})$ & $O(\eps^{-1})$ & $O(\eps^0)$
                  & $O(\eps^0)$    & $O(\eps^1)$ & $O(\eps^1)$ \\
$\ln\GammaN$
 & $O(\eps^{-2})$ & $O(\eps^{-2})$ & $O(\eps^{-1})$
                  & $O(\eps^{-1})$ & $O(\eps^0)$ & $O(\eps^0)$ \\ \hline
\end{tabular}
\end {center}
\caption
    {%
    \label {tablea}
    Calculations required for the $\eps$-expansion of various
    quantities discussed in the text.  C$_0$ indicates a tree-level
    calculation, and $C_n$ an $n$-loop calculation, at $\lambda(s)=0$;
    RG$_n$ indicates $n$-loop renormalization group flow.
    A computation through order $\eps^n$ requires all
    elements listed as $\eps^n$ in the table.
    }%
\end{table}

    As we have seen, various observables have different exponential
dependence on $1/\eps$ and $1/q^2$.
Consequently, determining their overall normalization can
require computing to different orders in perturbation theory.
Table \ref {tablea} summarizes the loop order needed for both the
renormalization group flow and the final calculation
in order to compute the observables we have discussed
to a given order in $\epsilon$.
One generally requires using renormalization group evolution
which is one higher order than the final calculation.
(This reflects the fact that
the renormalization group scale factor $\ln s$ is $O(1/\epsilon)$.)
The only exceptions are ratios, such as those in the second line of table
\ref {tablea}, in which the leading dependence on $s$ has been canceled.

\section {Two-loop analysis}

\subsection {Renormalization group flow}

The two-loop renormalization-group equations for the couplings have the form
\begin {mathletters}%
\begin {eqnarray}%
        s {d q^2\over ds}
    &=&
        \epsilon q^2 - \beta_0 \, q^4 - \beta_1 \, q^6 \,,
\\
        s {d\lambda \over ds}
    &=&
        \epsilon \lambda - (a \, q^4 + b \, q^2 \lambda + c \, \lambda^2)
        {} - (k_0 \, q^6 + k_1 \, q^4 \lambda
         + k_2 \, q^2 \lambda^2 + k_3 \, \lambda^3) \,.
\end {eqnarray}%
\label {srngg}%
\end {mathletters}%
We shall later be computing the latent heat and so, as in the one-loop
case, we also need the renormalization group equation for the mass,
\begin {equation}
        s {d m^2 \over ds}
    =
        -\left[
	    (l_0 \, q^2 + l_1 \lambda) +
	    (\ltwo_0 \, q^4 + \ltwo_1 \, q^2 \lambda + \ltwo_2 \, \lambda^2)
	\right] \, m^2 \,.
\end {equation}
The new constants above have the following values:
\vskip-\baselineskip
\begin {center}%
\begin {minipage}[b]{0.45\textwidth}%
\begin {eqnarray}%
{\rm U(1)}: \quad
  (4\pi)^4 \, \beta_1 &=& -8N                                  \,,\nonumber
\\    (4\pi)^4 \, k_0 &=& - 8 \, (45+7N)                       \,,\nonumber
\\    (4\pi)^4 \, k_1 &=&  {\textstyle {2 \over 3}}(87+71N)    \,,\nonumber
\\    (4\pi)^4 \, k_2 &=&  {\textstyle {8 \over 3}} (5+2N)     \,,\nonumber
\\    (4\pi)^4 \, k_3 &=& -{\textstyle {1 \over 3}}(14+6N)     \,,\nonumber
\\    (4\pi)^4 \, \ltwo_0 &=&  {\textstyle{1 \over3}}(15+71N)  \,,\nonumber
\\    (4\pi)^4 \, \ltwo_1 &=&  {\textstyle{16\over3}}(1+N)     \,,\nonumber
\\    (4\pi)^4 \, \ltwo_2 &=& -{\textstyle{5 \over9}}(1+N)     \,.\nonumber
\end {eqnarray}%
\end {minipage}%
\begin {minipage}[b]{0.55\textwidth}%
\begin {mathletters}%
\label {kcoeff}%
\begin {eqnarray}%
{\rm SU(2)}: \quad
  (4\pi)^4 \, \beta_1 &=& -{\textstyle {8 \over 3}}(544-13N)   \,,
\\    (4\pi)^4 \, k_0 &=& 168 \, (73-N)                        \,,
\\    (4\pi)^4 \, k_1 &=& -2 \, (455-71N)                      \,,
\\    (4\pi)^4 \, k_2 &=& 8 \, (5+2N)                          \,,
\\    (4\pi)^4 \, k_3 &=& -{\textstyle {1 \over 3}}(14+6N)     \,,
\\    (4\pi)^4 \, \ltwo_0 &=& -527+71N                         \,,
\\    (4\pi)^4 \, \ltwo_1 &=&  16 \, (1+N)                     \,,
\\    (4\pi)^4 \, \ltwo_2 &=& -{\textstyle{5 \over9}}(1+N)     \,.
\end {eqnarray}%
\end {mathletters}%
\end {minipage}%
\end {center}%
\vskip-\baselineskip
The constants for the flow of $\lambda$ and $q^2$ have been
extracted from the general results given in Ref.~\cite {two-loop-beta}.
As described in Appendix~\ref{beta-m appendix},
the $\beta$-function coefficients for the
mass have been extracted from the results of Ref.~\cite {Ford},
generalized to arbitrary $N$.

\begin {figure}
\vbox
    {%
    \begin {center}
	\leavevmode
	
	\epsfbox [72 310 500 460] {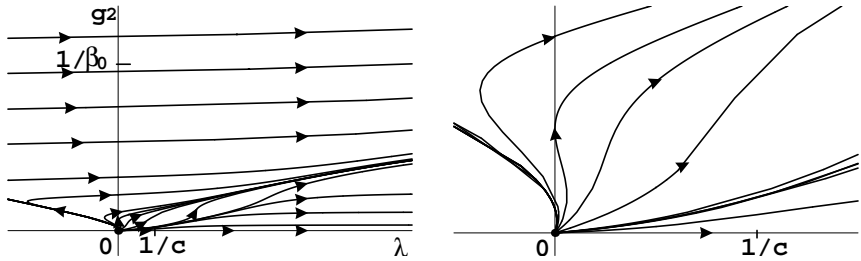}
    \end {center}
    \caption
	{%
	\label {figi}
	Two-loop renormalization group flows
	in three dimensions ($\epsilon=1$) for
	U(1) gauge theory with a single complex scalar.
	The right-hand figure shows an enlarged view
	of the region near the origin.
	}%
    }%
\end {figure}

In contrast to the one-loop renormalization group flows,
the two-loop trajectories are no longer independent
of $\epsilon$.
We will shortly derive the $O(\epsilon)$ correction to the one-loop solutions.
Since we are eventually going to take
$\epsilon \to 1$, it is tempting instead to simply plug
$\epsilon = 1$ into the equations (\ref {srngg}) and solve them numerically.
This is not the same as solving for the expansion in $\epsilon$,
truncating that expansion at next-to-leading order in $\epsilon$,
and then taking $\epsilon \to 1$.
It is also a temptation that should be resisted.
Fig.~\ref {figi} shows the result of solving
the two-loop equations directly in three dimensions ($\epsilon=1$)
for the U(1) gauge theory with $N=1$.
The trajectories are qualitatively very different than the
one-loop results of Fig.~\ref {figc}.
This might at first seem like a
disturbing problem for the $\epsilon$-expansion,
but similar behavior occurs even in pure scalar theory,
where the $\epsilon$-expansion is so successful.
The line $q^2=0$ of Fig.~\ref {figi} corresponds to
a complex pure scalar theory (or the $xy$ model).
The one-loop fixed point at $\lambda = 1/c$ has completely disappeared!
The two-loop equations with $\epsilon=1$ give no sign that the
phase transition in the scalar theory is second order!
This is equally true for the Ising model, which corresponds to $N=1/2$,
and whose critical exponents we reviewed in section \ref {scalarsec}.
The lesson is that one must keep strictly to the philosophy of
the $\epsilon$-expansion: take $\epsilon$ to be {\it small}, compute
results for {\it physical} quantities as a power series in
$\epsilon$, truncate such series at some finite order, and only
then send $\epsilon$ to 1.

\begin {figure}
\vbox
    {%
    \begin {center}
	\leavevmode
	
	\epsfbox [150 250 500 500] {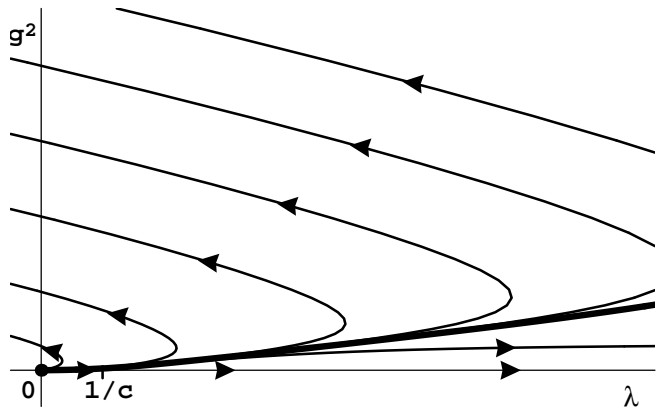}
    \end {center}
    \caption
	{%
	\label {figj}
	Two-loop renormalization group flows
	in three dimensions ($\epsilon=1$) for
	SU(2) gauge theory with a single scalar doublet.
        Flows below the heavy line run to $\lambda=+\infty$ instead of
        curving back to $\lambda = 0$.
	}%
    }%
\end {figure}

For the sake of comparison,
Fig.~\ref {figj} shows the analog of Fig.~\ref {figi}
for SU(2) theory with a single Higgs doublet ($N=2$).
The violence to the behavior of the one-loop flows is less severe,
but it is still qualitatively different.
Of course, the three-dimensional behavior of
the theory might really look something like Fig.~\ref {figj}
instead of Fig.~\ref {figc},
with some flows running to $\lambda = +\infty$;
but the goal of our current study is to investigate the
predictions of the $\epsilon$-expansion, in which case we must
follow the philosophy outlined above.

The $O(\epsilon)$ correction to $q^2(s)$ from the two-loop RG equation is
\begin {equation}
\label {eqggtwo}
   q^2(s) = \qa^2(s) + \qb^2(s) + O(\epsilon^3) \,,
\end {equation}
where $\qa^2$ is the one-loop solution
\begin {equation}
\label {gadef}
   \qa^2(s) = s^\epsilon q_1^2 \Bigm/
   \biggl[ 1 + {q_1^2\over q_0^2}(s^\epsilon {-} 1) \biggr] \,,
\end {equation}
and $\qb^2$ is the $O(\epsilon^2)$ correction
\begin {equation}
\label {gbdef}
   \qb^2(s) =
       - {\beta_1\over\beta_0} \, s^{-\epsilon} \, \qa^4(s) \left[
           (s^\epsilon {-} 1) - \left({q_0^2\over q_1^2}-1\right)
             \ln\left(s^\epsilon q_1^2\over \qa^2(s)\right)
         \right] .
\end {equation}
To find $\lambda(s)$, it is again convenient to solve the
renormalization group equation for the ratio $x=\lambda/q^2$,
\begin {equation}
\label {xrngtwo}
   s {d x\over ds} = - q^2 \, [a + (b-\beta_0)\, x + c \, x^2]
        - q^4 \, [ k_0 + (k_1-\beta_1) \, x + k_2 \, x^2 + k_3 \, x^3 ] ,
\end {equation}
whose solution is
\begin {equation}
\label {eqxxtwo}
   x(s) = \xa(s) + \xb(s) + O(\epsilon^2) ,
\end {equation}
where $\xa(s)$ is the one-loop result from (\ref {eqx}) and
\begin {eqnarray}
\label {eqxxb}
        \xb(s)
    &=&
        - e^{K(s)} \int_1^s {d s'\over s'} \> e^{-K(s')}
        \Bigl\{
            \qa^4(s') \,
        [k_0 + (k_1-\beta_1)\, \xa(s') + k_2\, \xa^2(s') + k_3\, \xa^3(s')]
\nonumber
\\
    &&
        \qquad \qquad \qquad \qquad \;
        {} + \qb^2(s') \, [a + (b-\beta_0)\, \xa(s') + c\, \xa^2(s')]
        \Bigr\} \,.
\end {eqnarray}
Here,
\begin {eqnarray}
\label {Kdef}
     K(s) &\equiv& \int_1^s {d s'\over s'} \>
                 \qa^2(s') \, \Bigl[ (\beta_0-b) - 2 c \, \xa(s') \Bigr]
\nonumber
\\
          &=& 2 \ln\sec \left[ \alpha - {\sqrt\Delta\over2\beta_0}
                          \ln\left(s^\epsilon q_1^2\over\qa^2(s)\right)
                       \right]
           - 2 \ln\sec \alpha \,,
\end {eqnarray}
with $\alpha$ and $\Delta$ given by Eqs.~(\ref {eqalpha}) and (\ref{eqDelta}),
respectively.
Solving $x(s)=0$ for $s$ now yields
\begin {equation}
   \label{seps2}
   s^\eps = \sa^\eps + \sb^\eps + O(\eps^2) \,,
\end {equation}
where $\sa$ is the one-loop result of (\ref{eqgs}) and
\begin {equation}
   \sb^\eps
       = - {\eps \, \sa^{\eps-1} \, \xb(\sa) \over
	   (\partial \xa(\sa) / \partial \sa)}
       = {\eps \, \sa^\eps \, \xb(\sa) \over a \, \qa^2(\sa)} \,.
\end {equation}
Substituting this result into Eq.~(\ref {eqggtwo}) yields the value
of $q^2(s)$ when $\lambda(s)$ vanishes,
\begin {eqnarray}
  \label{qq2}
q^2(s) &=&
   \qa^2(\sa)
   + \biggl({\sb^\eps \over \eps \, \sa^{\eps-1}} \,
		    {\partial \qa^2(\sa) \over \partial \sa}
                      + \qb^2(\sa)\biggr)
   + O(\eps^3)
\nonumber
\\
   &=&
   \qa^2(\sa)
   + \biggl( {\sb^\eps \over \eps \, \sa^\eps} \>
                      [\eps \, \qa^2(s) - \beta_0 \, \qa^4(\sa)]
            + \qb^2(\sa) \biggr)
   + O(\eps^3) .
\end {eqnarray}

\subsection {The two-loop potential}

The two-loop potential is one of the ingredients needed to compute
physical quantities, such as the correlation length,
beyond the leading order in $\eps$.
If the three-loop $\beta$-functions were known,
the two-loop potential could be used to obtain $\ln\xi\c$ to $O(\eps^1)$.
Without three-loop $\beta$-functions,
the two-loop potential still allows one to determine ratios such as
$\ln(\Delta Q/\xi\c^{2-\eps})$ to $O(\eps^1)$ as summarized in
Table~\ref{tablea}.

The two-loop potential at $\lambda(s)=0$ can be extracted in Landau gauge,
with small modifications, from the results of Ref.~\cite {Ford}.
In order to simplify the calculation, it is useful to note from
Eqs.~(\ref{eqvmc}) and (\ref{eqphic}) that within the $\epsilon$-expansion
the ratio of scalar to vector masses
is small near the asymmetric ground state at $\Tc$,
\begin {equation}
   {m^2 \over M^2} = {m^2 \over q^2 \bar\phi^{2^{\vphantom|}}} =
   O(q^2) = O(\eps) \,.
\end {equation}
Hence, if one neglects corrections suppressed by additional powers of $\eps$,
then the scalar mass may be set to zero when computing the two-loop potential.
The details of extracting the potential from Ref.~\cite {Ford},
as well as the result for general $\eps$, are given in
Appendix~\ref{potential appendix}.
Here, we shall simply quote the result when $\eps\to 0$,
\begin {equation}
   \mu^\eps V^{(2)}
   =
   \mu^\eps V^{(1)} +
   {q^6\bar\phi^4\over(4\pi)^4} \left[ \,
        v_2 \ln^2\left(q^2\bar\phi^2\over\bar\mu^2\right)
      + v_1 \ln\left(q^2\bar\phi^2\over\bar\mu^2\right)
      + v_0
   \right]
   \; (1+O(\eps)) \,.
   \label{VtwoEps}
\end {equation}
The coefficients $v_i$ are:
\vskip-\baselineskip
\begin {center}%
\begin {minipage}[b]{0.45\textwidth}%
\begin {eqnarray}%
{\rm U(1)}: \>
   v_2 &=&  {\textstyle {1\over4}} (9 {+} N)                    \,,\nonumber
\\ v_1 &=& -{\textstyle {1\over3}} (27 {+} 4N)                 	\,,\nonumber
\\ v_0 &=& {\textstyle {1\over12}}[123 {+} 19N - (1{-}N) \pi^2]	\,,\nonumber
\\ \phantom{{\pi\over6}}                          \nonumber
\end {eqnarray}%
\end {minipage}%
\hbox to -3.5em {\hss}%
\begin {minipage}[b]{0.63\textwidth}%
\begin {mathletters}%
\label {vcoeff}%
\begin {eqnarray}%
{\rm SU(2)}: \>
   v_2 &=& {\textstyle {3\over4}}  (-61 {+} N)			\,,
\\ v_1 &=& 286 {-} 4N						\,,\phantom{{1\over3}}
\\ v_0 &=& {\textstyle {1\over4}}[-19 {+} 19 N - (2 {-} N) \pi^2]
\\     &-& 297 \left[ \, 1 + \sqrt{3} \, L\biggl( {\pi \over 6} \biggr)
                   - \sqrt{3} \, {\pi\over6} \ln 2 \, \right]
		   \!,\!\!\!\!\!\!\!\!{} \nonumber
\end {eqnarray}%
\end {mathletters}%
\end {minipage}%
\end {center}%
\vskip-\baselineskip
where $L(t)$ is Lobachevskiy's function,
\begin {eqnarray}
   L(t) &\equiv& -\int\nolimits_0^t \d x \, \ln \, \cos x \,,
\label {lobachevsky}
\\
\noalign {\hbox {whose value at $\pi/6$ is}}
   L(\pi/6) &=& 0.024617 \,.
\end{eqnarray}

\subsection{The scalar correlation length}

As a sample computation using the two-loop potential,
we shall discuss the scalar correlation length $\xi\c$ at the critical
temperature.
At one-loop order, the correlation length is completely determined by
the curvature of the potential,
\begin {equation}
    \Omega\c^2 \equiv {\partial^2 \Veff \over \partial \phi^2} \,.
\end {equation}
At higher orders, one must be more careful to examine truly physical
quantities because the curvature of the potential is
both gauge- and scheme-dependent.
The physical correlation length $\xi$ is determined by the pole position
$p = i / \xi$ of the scalar propagator at zero frequency,
which satisfies the dispersion relation
\begin {equation}
   p^2 + m^2 + \Pi(p) = 0  \,,
\end {equation}
with $p_0 = 0$.
If one replaces $\Pi(p)$ by $\Pi(0)$, then the last two terms simply
produce the curvature of the effective potential.
The dispersion relation
may therefore be rewritten as
\begin {equation}
   p^2 + \Omega\c^2 + [\Pi(p) - \Pi(0)] = 0  \,.
   \label {dispersion}
\end {equation}
In the asymmetric state at $\Tc$, the curvature of the potential is
order $g^2 \mu^2$ while the momentum dependent part of the self energy,
$\Pi(p) - \Pi(0)$, is order $g^2 p^2 \mu/M \sim g^4 \mu^2$.
Locating the pole to leading order in $g$ then requires only the
one-loop effective potential, but at next-to-leading order one needs both
the two-loop potential and a one-loop calculation of $\Pi(p)-\Pi(0)$.

\begin {figure}
\vbox
    {%
    \begin {center}
	\leavevmode
	
	\epsfbox [150 320 500 460] {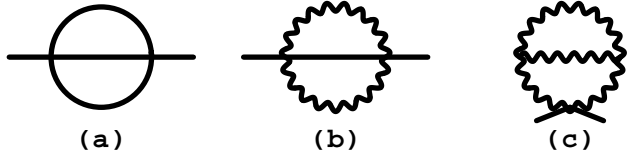}
    \end {center}
    \caption
	{%
	\label {figmdiv}
	Some two-loop contributions to the scalar mass that are logarithmically
	divergent in three dimensions.
	}%
    }%
\end {figure}

At one loop order we computed the scalar correlation length in both the
symmetric and asymmetric ground states.
At two loops, the symmetric correlation length becomes problematical in SU(2).
To see this, consider the two-loop potential directly in three dimensions
(using the $\eps\to 1$ limit of the results of
Appendix~\ref{potential appendix}).
It takes the form
\begin {equation}
   \mu V^{(2)}(\phi) \mathop\rightarrow\limits_{\epsilon=1}
       {\textstyle {1\over2}}
       \bar\phi^2
      \left( m^2 + {c_1 \, q^4\over 1{-}\eps}
                       + c_1 \, q^4 \ln\left(q^2\bar\phi^2\over\bar\mu^2\right)
                       + c_2 \, q^4 \right)
     - O(q^3) \, \mu \, \bar\phi^3
     + O(q^4) \, \bar\phi^4 \,,
\label {v3form}
\end {equation}
where $c_1$ and $c_2$ are constants.
We shall focus on the $\phi^2$ term.
The divergence as $\eps\to1$
is the usual logarithmic ultraviolet mass divergence arising at two loops
in three dimensions from graphs such as those illustrated
in Fig.~\ref{figmdiv}.
If we were doing perturbation theory directly in three dimensions,
we would adopt a renormalization scheme that absorbed this divergence
into the definition of a renormalized mass $m_{\rm ren}^2$;
physical quantities would then have a finite relationship to $m_{\rm ren}^2$
(rather than $m^2$) as $\eps\to 1$.
More troublesome than this ultraviolet divergence
is the infrared behavior of the graphs in Fig.~\ref{figmdiv},
which generate the $\phi^2 \ln M \sim \phi^2 \ln (q\phi)$
term in the potential (\ref{v3form}).
This term causes the curvature of the potential to diverge at $\phi=0$.
For graphs (a) and (b) of Fig.~\ref{figmdiv},
this divergence is an artifact of the
approximation where we neglected the scalar mass when evaluating the potential.
For the non-abelian graph (c), however, it is unavoidable.
As discussed in section \ref{magnetic mass section},
such infrared divergences are cut off by the
non-perturbative physics in the symmetric phase responsible for the
$O(q^2 \mu)$ mass gap for spatial gauge fields,
and this scale will cut off the logarithm in the two-loop correction
to the scalar correlation length.
This infrared cut-off of the $\phi^2 \ln (q\phi)$ term at scale where
$q\phi \sim q^2\mu$
also remedies a potentially embarrassing feature of the perturbative result
(\ref{v3form})
when $c_1$ is positive (which occurs for sufficiently large $N > 19$):
$\phi=0$ is then a local {\it maximum} instead of minimum,
with a nearby local minimum at $q \phi \sim \mu \exp[-O(1/q^2)]$.

    In the symmetric phase, the two-loop correlation length has a
direct dependence on non-perturbative physics because this
is the only source of an infrared cut-off.
In the asymmetric phase, however,
there is already a cut-off from the non-zero value of $q\phi$.
We shall therefore focus exclusively on the asymmetric phase,
where the two-loop correction to the correlation length remains sensible
even when $\eps \to 1$.
It would be interesting to compute multi-loop corrections to the
symmetric phase correlation length within the $\epsilon$-expansion
and investigate the extrapolation to $\epsilon = 1$,
but we have not done so.

    To find the asymmetric phase correlation length,
the first task is to compute the curvature of the two-loop potential
at the asymmetric minimum $\phi\c$.
Perturb around the one-loop solution by writing $V = V^{(1)} + \delta V$,
$\phi\c = \phi\c^{(1)} + \delta\phi\c$,
and $m\c^2 = [m\c^{(1)}]^2 + \delta m^2$,
and then linearize
the equations $V(\phi\c) = V(0)$ and $(\partial / \partial \phi) V(\phi\c) = 0$
that determine the asymmetric state at the critical temperature.
Solving for $\phi\c$ and $m\c^2$,
one finds that
\begin {eqnarray}
	\delta\bar\phi\c
    &=&
	-{4! \> \mu^\eps \over 2 a \, q^4} \,
	{\partial \over \partial \bar\phi}
	\left(
	    {\delta V(\bar\phi) - \delta V(0) \over \bar\phi^2}
	\right)
	\Biggr|_{\bar\phi = \bar\phi\c^{(1)}} \,,
\\
	\delta m^2
    &=&
	-2 \mu^\eps
	\left(
	    {\delta V(\bar\phi) - \delta V(0) \over \bar\phi^2}
	\right)
	\Biggr|_{\bar\phi = \bar\phi\c^{(1)}} \,,
\end {eqnarray}
and the curvature of the potential at $\phic$
differs from its one-loop value by
\begin {equation}
    \delta \Omega\asym^2 =
    \Bigl(
            \bar\phi^2 {\partial^2 \over \partial \bar\phi^2}
             - (3 {-} \eps) \, \bar\phi {\partial \over \partial \bar\phi}
    \Bigr)
    \left(
        \delta V(\bar\phi) - \delta V(0) \over \bar\phi^2
    \right)
    \Biggr|_{\bar\phi = \bar\phi\c^{(1)}} \,.
\end {equation}
Plugging in the general results for $\delta V$
from Appendix~\ref{potential appendix}
yields
\begin {equation}
   \Omega\asym^2 =
       {T^2\over s^2} f(\eps)^2 q^2(s) {a\over 4!} \,
       {8\pi\over e^\gammaE}
       \left[ 1 + {q^2(s)\over(4\pi)^2} \, S(\eps) + O(q^4(s))\right] \,,
\label {curv-result}
\end {equation}
where
\begin {eqnarray}
   S_{\rm U(1)}(\eps) &=&
     {\textstyle {4\over3}}
     \Biggl\{
        {1 \over \epsilon^2} \,
        { \left(1+{\eps\over2}\right) \left(1-{\eps\over4}\right)^2 \over
          \left(1-{\eps\over2}\right) \left(1-{\eps\over3}\right)^2 }
       \left[ (N {-} 1) \eps B\left(-{\eps / 2},\eps\right)
              - {4 \, (5-4\eps+\eps^2) \over 2-\eps} \right]
\nonumber
\\ && \qquad \qquad \qquad \qquad \qquad
       {} + {9+N\over\eps^2} + {45+7N\over6\eps}\left(1-{\eps\over2}\right)
      \Biggr\} \,,
\label{OmegaU1}
\\
   S_{\rm SU(2)}(\eps) &=&
     {\textstyle {4\over3}}
     \Biggl\{^{\phantom {\strut}}
            {1 \over \epsilon^2} \,
            { \left(1+{\eps\over2}\right) \left(1-{\eps\over4}\right)^2 \over
              \left(1-{\eps\over2}\right) \left(1-{\eps\over3}\right)^2 }
           \Biggl[ \left( -2 + N - (11 {-} 4\eps) \, 3^{(3-\eps)/2}
                     \sin\left(\eps\pi / 2\right) \right)
               \eps B\left(-{\eps / 2},\eps\right)
\nonumber
\\ && \qquad\qquad\qquad\qquad\qquad
              {} - {4 \, (20-35\eps+21\eps^2-4\eps^3) \over 2-\eps}
              + 9 \, (11 {-} 4\eps)(1 {-} \eps) \, r(\eps)
           \Biggr]
\nonumber
\\ && \qquad\qquad\qquad\qquad\qquad
       {} - {61-N\over\eps^2}
       - {7 \, (73-N)\over6\eps}\left(1-{\eps\over2}\right)
      \Biggr\} \,.
\label{OmegaSU2}
\end {eqnarray}
Here,
$B(\mu,\nu)$ is Euler's Beta function, and the function
$r(\eps)$ is defined in Eq.~(\ref {rdef}) of the appendix.
These results are indeed smooth as $\eps\to1$.
The result for small $\eps$ (which can be derived directly from the
the $\eps\to 0$ limit of Eq.~(\ref{VtwoEps})) is
\begin {equation}
   \label{OmegaEps}
   S(\eps) = {1\over\ng} \left(
                 {\textstyle{71\over27}} v_2
               + {\textstyle{2\over9}} v_1
               - {\textstyle{4\over3}} v_0
             \right) \> [1+O(\eps)] \,.
\end {equation}

\begin {figure}
\vbox
    {%
    \begin {center}
	\leavevmode
	
	\epsfbox [150 330 500 420] {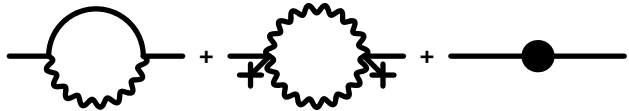}
    \end {center}
    \caption
	{%
	\label {fig:Pi-diff}
	One-loop graphs contributing to $\Pi(p)-\Pi(0)$.  The black dot
	in the last graphs represents the one-loop counter term for
	wave-function renormalization.
	}%
    }%
\end {figure}

To complete the calculation of the correlation length, we need
a one-loop calculation of $\Pi(p)-\Pi(0)$ to order $q^4 \mu^2$.
The graphs which contribute to $\Pi(p)-\Pi(0)$ are shown
in fig.~\ref{fig:Pi-diff}.  The calculation of these graphs may
be simplified by remembering that $m^2/M^2 \sim q^2 \sim \eps$
at $\Tc$.  So $m^2$ and $p^2$ may be treated as small perturbations
compared to $M^2$ when evaluating these graphs.  In Landau gauge,
one finds
\begin {eqnarray}
   \Pi(p) - \Pi(0)
   &\equiv& {q^2 p^2 \over (4\pi)^2} \, \bar S(\eps)
\label {mom-dep}
\end {eqnarray}
with
\begin {eqnarray}
       \bar S(\eps)
   &=&
   \ng \left[
       {6\over\eps}
       + \left(4\pi\mu^2\over M^2\right)^{\eps/2}
             (3 {-} \eps) \left(1 - {\eps\over6} + {\eps^2\over12}\right)
             \Gamma \!\left(-1 {+} {\eps\over2}\right)
   \right] \>
   \left[ 1 + O \! \left({m^2\over M^2}, {p^2\over M^2}\right)\right]
\\
   &=&
   -3 \ng
       \ln\left(\bar\mu^2\over M^2\right)
       ~\left[ 1 + O \! \left(q^2, {q^2 p^2\over m^2}\right) \right] \,.
\end {eqnarray}
Plugging in the value of $M = q\phi$ using (\ref{eqphic}) gives
\begin {eqnarray}
   \bar S(\eps) &=&
   {6\ng\over\eps} \left[ 1 -
      \left(1-{\eps\over2}\right)^{-1}
      \left(1-{\eps\over4}\right)	
      \left(1-{\eps\over6}+{\eps^2\over12}\right)
   \right]
   \left[ 1 + O \!\left(q^2, {q^2 p^2\over m^2}\right)\right]
\\
   &=&
   - {\textstyle{1\over2}} \ng
   ~\left[ 1 + O\! \left(\eps, \eps {p^2\over m^2}\right) \right] \,.
   \label {SbarEps}
\end {eqnarray}
Inserting results (\ref {curv-result}) and (\ref {mom-dep}) into
the dispersion relation (\ref{dispersion}), the final result
for the correlation length in the asymmetric phase is
\begin {equation}
   \label{xiasym2}
   \xi\asym = {s\over T} \, (f(\eps)q(s))^{-1}
        \sqrt{{4!\over2a} {e^\gammaE\over4\pi}}
       \left[
	   1 + {\textstyle {1\over2}} \, { q^2(s)\over(4\pi)^2}
	       \left( \bar S(\eps) -  S(\eps) \right) + O(q^4)
       \right] \,.
\end {equation}

\subsection {Testing the $\eps$-expansion when $\lambda_1 \ll q_1^2$}

As we did for the one-loop case, it is interesting to compare our
two-loop corrections to the correct three-dimensional results in the
perturbative regime $\lambda_1 \ll q_1^2$.  We first need the
expansions of $s^\eps$ and $q^2(s)$ in powers of $\lambda_1/q_1^2$
at $\lambda(s) = 0$.
The expansion of $s^\eps$ can be obtained with some sweat by expanding the
two-loop result (\ref{eqxxtwo}) for $x(s)$ and setting $x(s) = 0$.
It is much easier, however, to return to the original two-loop
RG equations (\ref{srng}) and solve them perturbatively in
$x=\lambda/q^2$, recalling from the one-loop analysis (\ref {george})
that $q^2(s) \sim s^\eps q_1^2 \sim \eps x_1$.
One finds,
\begin {eqnarray}
   q^2(s) &=& s^\eps q_1^2 - s^\eps (s^\eps {-} 1) \,
   { q_1^4\over \epsilon} \, \beta_0
            + O(\eps x_1^3) \,,
\\
   x(s) &=& x_1
       - (s^\eps {-} 1) \, {q_1^2 \over \eps} \, (a + (b {-}  \beta_0)\, x_1)
	      + (s^\eps{-} 1)^2 {q_1^4 \over 2 \eps^2} \, a b
	      - (s^{2 \eps} {-} 1) \, {q_1^4 \over 2 \eps} \, k_0
\nonumber
\\ && \qquad {}
	      + O(x_1^2 \, q^2(s) \ln s) \,.
\end{eqnarray}
Setting $x(s)$ to zero and, as usual, assuming $q_1^2$ is small
by taking the limit $q_1^2 \to 0$ with $x_1$ fixed yields
\begin {eqnarray}
   q^2(s) &=& {\eps x_1 \over a}
		\left (
		    1
		    - {x_1 \over 2a} \bigg(b + {\eps k_0\over a}\bigg)
		    + O(x_1^2)
		\right) ,
\\
   s^\eps &=& {\eps x_1 \over a q_1^2}
		\left(
		    1
		    + {x_1 \over 2a} \bigg(2\beta_0 -b - {\eps k_0\over a}\bigg)
		    + O (x_1^2)
		\right) .
\end {eqnarray}
Applying this expansion to our results (\ref{OmegaU1}) and (\ref{OmegaSU2})
for the two-loop curvature yields
\begin {eqnarray}
   \xi\asym &=&
   {1\over T} \sqrt{q_1^2\over\eps\lambda_1}
   \left( \eps \lambda_1 \over a q_1^4 \right)^{1/\eps}
   \bigl[
      M_0(\eps) + M_1(\eps) \, x_1 + O(x_1^2)
   \bigr] \,,
\end {eqnarray}
where $M_0(\eps) = 1.41748 - 0.25701 \eps + \cdots$ is the previous series
in Eq.~(\ref {eqexs}), and
\begin {eqnarray}
M_1^{\rm U(1)}(\eps) &\equiv& {}
      (0.23625+0.02625 \,N) \, \eps^{-1}
      + (0.03591+0.02587 \,N)
\nonumber \\ && {}
      + (0.06521+0.03711 \,N) \, \eps
      + (0.01481+0.02248 \,N) \, \eps^2
\nonumber \\ && {}
      + (0.01249+0.02594 \,N) \, \eps^3
      + (0.01041+0.00964 \,N) \, \eps^4
\nonumber \\ && {}
      + ({-}0.00035+0.01468 \,N) \, \eps^5
      + O(\eps^6) \,,
\label {eqexsa}
\\
M_1^{\rm SU(2)}(\eps) &\equiv& {} \;
      ({-}0.53374+0.00875 \,N) \, \eps^{-1}
      + ({-}0.76655+0.00862 \,N)
\nonumber \\ && {}
      + (0.21722+0.01237 \,N) \, \eps
      + ({-}0.26359+0.00749 \,N) \, \eps^2
\nonumber \\ && {}
      + (0.21748+0.00865 \,N) \, \eps^3
      + ({-}0.22786+0.00321 \,N) \, \eps^4
\nonumber \\ && {}
      + (0.2219\  +0.00489 \,N) \, \eps^5
      + O(\eps^6)  \,.
\label {eqexsb}
\end {eqnarray}
The correct, three-dimensional result is
\begin {equation}
M_1^{\rm U(1)}(1) = 0.38119 + 0.17444 N \,,
  \qquad
M_1^{\rm SU(2)}(1) = -1.24695 + 0.05815 N \,.
\end {equation}
The convergence of the $\epsilon$-expansion for the $O(x_1)$
terms seems somewhat worse for the SU(2) case than the leading
$(x_1)^0$ terms that we examined previously.%
\footnote
    {%
    This can be traced to the pole of the Beta function in
    (\ref{OmegaSU2}) at $\eps=-1$, which makes the radius of
    convergence equal to 1.  This pole reflects the appearance
    of new, logarithmic two-loop divergences in five dimensions unrelated
    to those in four dimensions.
    }
Nevertheless, in the SU(2) case with $N=2$,
when $\eps\to 1$ the terms through $O(\eps^0)$
in the coefficient $M_1(\eps)$ of $x_1$
give a result differing from the three-dimensional result by 12\%,
and the sum of terms through $O(\eps^1)$ err by only $-9$\%.
After that, higher-order terms give larger and larger contributions,
which is the signal to stop.
So, either the $O(\eps^0)$ or $O(\eps^1)$ results give
reasonable approximations to the correct answer.
Keep in mind that, since we have assumed $x_1$ is small,
we are currently testing the $\eps$-expansion on a small
{\it correction} to the total result.

\subsection {The latent heat as a test for $\lambda_1/q_1^2 \gsim 1$.}
\label {2 loop latent sec}

We shall now test the $\eps$-expansion in the range
$\lambda_1/q_1^2 \gsim 1$, where unimproved perturbation theory
fails, by computing a sample quantity to next-to-leading order
and checking whether the correction to the leading-order result
is large or small.  Recall from the discussion in section
\ref{applications1-sec} that, without three-loop RG equations,
we cannot consistently compute next-to-leading order corrections
to the prefactors of quantities such as the correlation length or
the latent heat.  But we can compute the correction to
the ratios such as
$\xi\asym^2 \Delta Q$ with a purely two-loop calculation using the
two-loop renormalization group.
In this section, we do a next-to-leading order calculation
of $\xi\asym^2 \Delta Q$ to test the behavior of the $\eps$-expansion.

We first need to compute $\Delta Q$ by using (\ref{DQrelation}) and
linearizing about the one-loop result.  We won't derive the
two-loop result for arbitrary dimension but shall instead simply
use the $\eps\to0$ limit (\ref{VtwoEps}) for the two-loop potential.
One finds
\begin {eqnarray}
\label {moo2}
	{\Delta Q\over \nu^2}
   &=&
	{T^{3-\eps} \over s^{2 - \eps}} \,
	{f(\eps)^2 \over q^2(s)} \,
	{4\pi\over e^\gammaE} \,
	P(s^\eps)
        \left\{ 1 + {q^2(s)\over(4\pi)^2} {1\over\ng}\left(
                       \textstyle{11\over27} v_2
                     - \textstyle{10\over9} v_1
                     - \textstyle{4\over3} v_0 \right)
               + O(\eps^2)
        \right\} \,,
\end {eqnarray}
and $\xi\asym^2 \Delta Q$ follows from the results
(\ref{OmegaEps}), (\ref{SbarEps}) and (\ref{xiasym2}) for
$\xi\asym$,
\begin {eqnarray}
\label {xiQratio2}
       {\xi\asym^2\Delta Q\over \nu^2}
   &=& {T^{1-\eps} \over s^{-\eps}} \,
       {4! \over 2a} \, {P(s^\eps) \over q^4(s)}
        \left\{ 1 - {q^2(s)\over(4\pi)^2}
	    \left[
		  {1 \over \ng}
		  \left(
		      {\textstyle {4\over3}} v_1 + {\textstyle {20\over9}} v_2
		  \right)
		  + {\textstyle {1\over2}} \ng
	    \right] + O(\eps^2)
	\right\} .
\end {eqnarray}
Here,
$P(s^\eps)$ is the two loop version of the
ratio (\ref{Pdef}) of the running mass at the initial and final scales,
\newpage
\begin{eqnarray}
  \label {Pdef2}
   P(s^\eps)
   &\equiv& {m^2 (\mu) \over m^2_1}
   = \exp\biggl\{-{1\over\eps} \int\nolimits_1^{s^\eps}
          {ds'^\eps \over s'^\eps} \>
          \Bigl[ l_0 \, q^2(s') + l_1 \, \lambda(s')
\nonumber
\\ && \qquad\qquad\qquad\qquad\qquad\qquad {}
                 + \ltwo_0 \, q^4(s') + \ltwo_1 \, q^2(s') \lambda(s')
                 + \ltwo_2 \, \lambda^2(s')
          \Bigr] \biggr\}
\nonumber
\\ &=& P_a (s_a^\eps) \> [1 + \delta_1 + O(\eps^2)] ,
\\
\noalign{\hbox {where $P_a$ is the one-loop mass ratio,}}
    P_a (s_a^\eps) &\equiv&
    \exp\biggl\{-{1\over\eps} \int\nolimits_1^{\sa^\eps}
          {ds'^\eps \over s'^\eps} \>
          \qa^2(s') \, [ l_0 + l_1 \, \xa(s') ]
       \biggr\} \,,
\label {P_a}
\\
\noalign{\hbox {and $\delta_1$ is the two-loop correction,}}
   \delta_1 &\equiv&
   - {\sb^\eps \over \eps \, \sa^\eps} \, l_0 \, \qa^2(\sa)
   - {1\over\eps} \int\nolimits_1^{\sa^\eps}
      {ds'^\eps \over s'^\eps} \>
      \biggl\{
         l_0 \, \qb^2(s') + l_1 \, [ \qa^2(s') \xb(s') + \xa(s') \qb^2(s') ]
\nonumber
\\ && \qquad\qquad\qquad\qquad\qquad\qquad {}
         + \qa^4(s') \, [\ltwo_0 + \ltwo_1 \,\xa(s') + \ltwo_2 \,\xa^2(s')]
      \biggr\} \,.
\end {eqnarray}
Finally, we need the expansion (\ref {qq2}) of 
$q^2(s)$ at $\lambda(s)=0$, or
\begin {eqnarray}
   \label {qm4exp}
   {1\over q^4(s)} &=& {1\over\qa^4(\sa)} \> (1 - 2 \delta_2 + O(\eps^2)) ,
\\
\noalign{\hbox{where}}
   \delta_2 &=&
      {\sb^\eps \over \eps \,\sa^\eps} \left(\eps - \beta_0 \,\qa^2(\sa)
\right)
      + {\qb^2(\sa)\over\qa^2(\sa)} \,.
\end {eqnarray}
Inserting these expansions of $q^2(s)$ and $P(s^\eps)$,
plus that of the scale $s^\eps$, in Eq.~(\ref {seps2}),
into the result (\ref {xiQratio2}) yields
\begin {eqnarray}
       {\xi\asym^2\Delta Q\over \nu^2}
   &=& {T^{1-\eps} \over \sa^{-\eps}} \, {4! \over 2a} \,
       {P_a (\sa^\eps) \over \qa^4(s)} \;
        [ 1 + \delta_{\rm tot} + O(\eps^2) ] \,,
\\
\noalign{\hbox{where}}
   \delta_{\rm tot} &=&
       \delta_1 - 2\delta_2 + {\sb^\eps\over\sa^\eps}
                - {\qa^2(\sa)\over(4\pi)^2}
	    \left[
		  {1 \over \ng}
		  \left(
		      {\textstyle {4\over3}} v_1 + {\textstyle {20\over9}} v_2
		  \right)
		  + {\textstyle {1\over2}} \ng
	    \right] \,.
   \label {deltoteq}
\end {eqnarray}

\begin {figure}
\vbox
    {%
    \begin {center}
	\leavevmode
	
	\epsfbox [150 250 500 500] {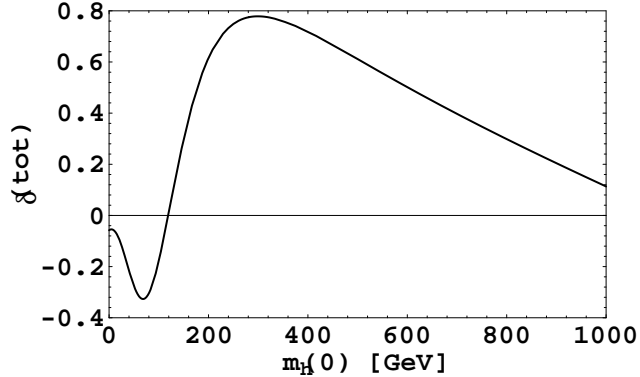}
    \end {center}
    \caption
	{%
	\label {figlat}
        The relative size of the next-to-leading order correction
        to $\xi\asym^2 \Delta Q$ in the $\eps$-expansion.  The values
        are given as a function of the (tree-level)
        zero-temperature Higgs mass in minimal SU(2)
        theory ($N=2$) with $g = 0.63$.
	}%
    }%
\end {figure}

The relative size of this $O(\eps)$ correction is shown in
Fig.~\ref{figlat} for the minimal standard model (in our approximation
of ignoring the Weinberg mixing angle).
The correction varies between roughly $\pm 30$\% for
(zero-temperature) Higgs masses up to 150 GeV.
This suggests that the $\eps$-expansion is tolerably well-behaved
for these masses.
For larger masses, the correction does not increase indefinitely
but is bounded by 80\%.
This gives hope that the $\eps$-expansion may be a useful
qualitative description of the phase transition even for these
larger masses where it does not work as well quantitatively.

The comparatively small size of $\delta_{\rm tot}$
becomes more impressive when one
examines the size of the four separate terms in (\ref{deltoteq}).
At $\mh(0) = 80$ and 250 GeV, they are
\begin {eqnarray}
   \delta_{\rm tot}(\phantom{2} 80~\hbox{GeV})
      &=& -0.45 + \phantom{3} 4.88 - \phantom{2} 0.94 - 3.79
      = -0.30 \,,
\\
   \delta_{\rm tot}(250~\hbox{GeV})
      &=& -4.42 + 34.83 - 20.00 - 9.66
      = \phantom{-} 0.75 \,.
\end {eqnarray}
These large cancellations clearly underscore the importance of examining
{\it physical} quantities, rather than unphysical ones such as
$q^2(s)$ or $s^\eps$, when testing the $\eps$-expansion.

\section {Many scalar fields}

In this section, we examine two calculations related to
theories where the number of complex scalar fields, $N$, is large.
As we shall see, the $\eps$-expansion is less well behaved in such theories.

\subsection {Critical $\Nc$}

As noted in the introduction, a non-trivial infrared stable fixed point
exists in both U(1) and SU(2) theories if the number of scalar fields
is sufficiently large, $N \ge N_c$.
As $\epsilon\to 0$, there can never be an infrared-stable fixed point
if the gauge coupling $q^2$ is asymptotically free,
since then the one-loop renormalization group equation
(\ref {srngA}) has no non-trivial zero.
A critical number $\Nc$ of scalar fields exists for SU(2)
only because $\beta_0$ becomes positive for $N > 88$,
thereby permitting a non-trivial fixed-point for sufficiently large $N$.
If there is a fixed point, it must have
\begin {equation}
\label {gstar}
   q_*^2 = {\epsilon\over\beta_0} + O(\epsilon^2) .
\end {equation}
A fixed point must also have $d x/d s = 0$, where
this derivative can be obtained from Eq.~(\ref {xrngtwo}).
As $N$ approaches $\Nc$ from above, the stable fixed point $A$ and
the tricritical point $B$ shown in Fig.~\ref {fige} merge.
This provides a simple way to compute $\Nc$, for when the
two fixed points touch not only must the right-hand side of the
RG equation (\ref {xrngtwo}) vanish,
so must its derivative with respect to $x$.
In particular, the combination $(1- {1 \over 2} x\partial / \partial x)$
must give zero when applied to the right-hand side, which implies that
\begin {equation}
\label {eqncxx}
   \left( a + {\textstyle {1\over2}} (b-\beta_0) \, x_* \right)
   + q_*^2
   \left(
       k_0 + {\textstyle {1\over2}} (k_1-\beta_1) x_*
           - {\textstyle {1\over2}} k_3 \, x_*^3
   \right)
   = 0 \,.
\end {equation}
This has the solution
\begin {equation}
\label {eqncx}
   x_* = {2a\over\beta_0-b} +
       {2 q_*^2\over\beta_0-b} \left( k_0 + {a \, (k_1-\beta_1)\over\beta_0-b}
              - {4 k_3 \, a^3\over(\beta_0-b)^3} \right)
       + O(\epsilon^2) .
\end {equation}
Plugging back into the RG equation $d x/d s = 0$ gives
\begin {equation}
\label {eqnc}
   \Nc = \Nc^{(0)} + \epsilon \, \Nc^{(1)} + O(\epsilon^2)
\end {equation}
where $\Nc^{(0)}$ is the value of $N$ for which
$\Delta = 4 a c - (b-\beta_0)^2$ is zero and $\Nc^{(1)}$ is
\begin {equation}
        \Nc^{(1)} =
        - {4 / \beta_0 \over (\partial \Delta / \partial N) }
        \left[ \,
            c \, k_0
            + {\textstyle {1\over2}} (\beta_0-b) (k_1-\beta_1)
            + a k_2
            + 2 a^2 \, k_3 / (\beta_0-b)
        \right] \biggr|_{N=\Nc^{(0)}} .
\label {eqncb}
\end {equation}
Inserting the coefficients from (\ref {kcoeff}) yields the results
\begin {eqnarray}
   \Nc^{\rm U(1)}(4{-}\epsilon) &=& 182.95 - 242.67 \, \epsilon + O(\epsilon^2)
   \,,
\label {ncabel}
\\
   \Nc^{\rm SU(2)}(4{-}\epsilon) &=& 718 - 990.83 \, \epsilon + O(\epsilon^2) .
\label {ncsu}
\end {eqnarray}
As noted in the introduction, these series clearly  can not be used to
obtain a reliable estimate of $\Nc$ when $\epsilon = 1$.

\subsection {Tricritical slope \lowercase {$\lambda/q^2$}}

The goal of this section is to compute,
for large $N$,
the ratio $\lambda/q^2$ of the tricritical point separating
the domains of first and second order phase transitions,
as shown in Fig.~\ref {fige}.
The lowest-order term in the $\epsilon$-expansion
is given by the $N \to \infty$ limits of the expression
(\ref {eqxpm}) for $x_-$.
The first order correction is easily determined
by solving for the zero of the two loop renormalization group
equation (\ref {xrngtwo}) to order $\epsilon$.
One finds
\begin {equation}
\label {eqxcr}
   {\lambda\over q^2} =
   \left[ \, 54 -126 \, \eps + O(\eps^2)\right] \, {\ng \over N}
		   + O\left({1 \over N^2}\right) \,.
\end {equation}
This $\eps$-expansion is also poorly behaved.
We shall see below that, though the qualitative dependence on
$N$ is correct, the leading-$O(\epsilon^0)$ coefficient of $54$
differs from the correct three-dimensional coefficient by a
factor of roughly five!  It is important to emphasize, however,
that the $\eps$-expansion alerts one to its own failure
by producing a next-to-leading order correction that is
significantly larger than the leading-order result when $\epsilon=1$.

Even the qualitative result that the critical $\lambda/q^2$
scales like $1/N$ is at odds with a large $N$ analysis in three
dimensions by Jain and Papadopoulos \cite{Jain}.
One of the main purposes of this section is to redo their
analysis and show that the $\eps$-expansion result is
qualitatively correct.  We shall closely follow their
analysis, though our final result is different.
It should be emphasized that large $N$
in the present context means a large number of scalars,
not the replacement of the gauge group by SU($N$).

Before focusing on three dimensions, we note a qualitative
feature of the $\eps$-expansion result which will help guide
us in selecting the orders of parameters of interest in the
three dimensional case.  The tricritical point determined by
(\ref{eqxcr}) approaches the $\lambda = 0$ axis as $N \to \infty$.
Recalling that the value of $q^2$ at the tricritical point scales
like $1/N$, the prediction for $\lambda$ at the tricritical point
is $O(1/N^2)$.

Now consider the three dimensional theory
(with either U(1) or SU(2) gauge group).
Taking a hint from the $\epsilon$-expansion, examine the regime
where $q^2$ is order $1/N$ and $\lambda$ is order $1/N^2$.
We want to compute the effective potential in the
large-$N$ limit and determine whether its variation with $m^2$
yields a first- or second-order phase transition.
Since the effective potential is a function of $\phi$,
evaluating its large-$N$ limit depends on how 
$\phi$ scales with $N$.
To get a hint, recall the results of naive one-loop perturbation theory.
This produces a potential of the form
\begin {equation}
\label {eqvna}
  V(\phi) \sim {\textstyle {1\over2}} m^2 \phi^2 -
                \kappa \, q^3 \mu^{3/2} \phi^3 +
                {\textstyle {1\over4!}} \, \lambda \mu \,\phi^4 \,,
\end {equation}
where the constant $\kappa$ is $O(1)$.
In the asymmetric phase at the transition,
all three terms are of the same order, which implies that
\begin {mathletters}
\begin {eqnarray}
\label {eqnord}
    \phi &\sim& {q^3 \sqrt \mu \over \lambda} = O(\sqrt{N}),
\\
    m^2 &\sim& \lambda \, \mu \, \phi^2 = O\left(1 / N\right) \,.
\end {eqnarray}
\end {mathletters}%
In this range of $\phi$ the vector mass $M$
is large compared to $m$,
\begin {equation}
\label {eqnordM}
    M^2 = q^2 \mu \phi^2 = O(N^0) \, .
\end {equation}

\begin {figure}
\vbox
    {%
    \begin {center}
	\leavevmode
	
	\epsfbox [150 270 500 500] {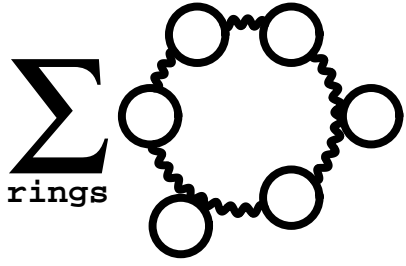}
    \end {center}
    \caption
	{%
	\label {figaa}
	Dominant graphs in large $N$ for $\lambda/q^2 \sim 1/N$.
	}%
    }%
\end {figure}

Using these parameters, we may easily compute the three-dimensional
effective potential to leading order in $N$.
The dominant graphs are shown schematically
in Fig.~\ref {figaa} and are vector loops decorated by scalar rings.
Any
other type of graph is sub-leading in $N$.
Hence, in Landau gauge,
the leading-order potential (neglecting $\phi$-independent terms)
is simply
\begin {equation}
\label {eqnpot}
    V(\phi) = {\textstyle {1\over2}} m^2\phi^2 +
              {\textstyle {1\over4!}} \lambda\mu\phi^4 +
               \ng \int{d^3p\over(2\pi)^3} \> \ln(p^2+M^2+\Piv(p)) \,.
\end {equation}
Here,
$\Piv(p)$ is the scalar contribution to the one-loop vector self-energy.
The dominant momenta flowing through the ring graphs are of order $M$.
Consequently, in the self energy $\Piv(p)$ the scalar mass $m^2$
is small compared to $p^2$ and may be neglected.
The leading contribution to $\Piv(p)$ is then easily computed to be
\begin {equation}
\label {eqnpi}
  \Piv(p) = {\textstyle{1\over16}} N q^2 p \mu \,.
\end {equation}

\begin {figure}
\vbox
    {%
    \begin {center}
	\leavevmode
	
	\epsfbox [150 270 500 530] {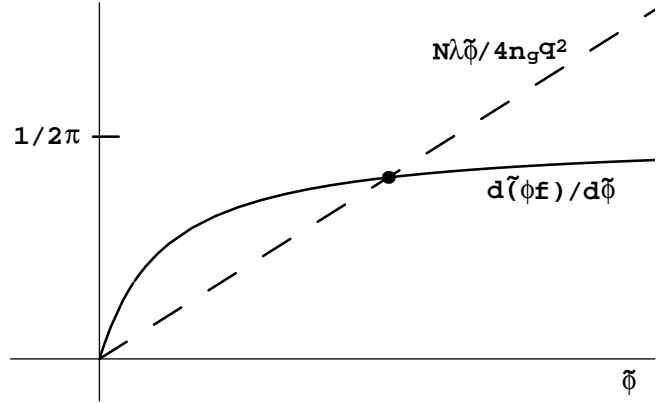}
    \end {center}
    \caption
	{%
	\label {figab}
	Graphical representation of solution to
	Eq.~(\protect {\ref {eqnfconst}}).
	}%
    }%
\end {figure}

We now need to evaluate the integral in (\ref {eqnpot}).
This integral must first be regulated and renormalized.
Two subtractions are needed to render the integral finite,
one $\phi$-independent and one proportional to $\phi^2$.
The resulting finite, renormalized integral is
\begin {equation}
\label {eqnlog}
  \int {d^3p \over (2\pi)^3} \>
    \left\{
	\ln \left({p^2+M^2+\Piv(p) \over p^2+\Piv(p)} \right)
	-{M^2 \over p^2+\Piv(p)}
    \right\}
  =
  - {\textstyle {1\over3}} \, M^3 f(z),
\end {equation}
where
\begin {equation}
\label {eqnf}
  f(z) \equiv {z^{-3/2} \over \pi^2}
      \left[
          (z {-} 4)\sqrt{z {-} 1} \, {\rm Sec}^{-1}\sqrt{z}
          - \left( {3\over2} \, z - 2\right)
            \ln\left( z\over4 \right)
          + z \,
      \right] \,,
\end {equation}
and
\begin {equation}
\label {nzdef}
   z \equiv {1024 \, M^2 \over N^2 q^4 \mu^2}
   = {1024 \, \phi^2 \over N^2 q^2 \mu} \,.
\end {equation}
The resulting potential can then
be written in the form
\begin {equation}
\label {eqnv}
  V(\tilde\phi) = (N q^2 \mu)^3
  \left[
             {\textstyle {1\over2}} \tilde m^2 \tilde\phi^2
           - {\textstyle {1\over3}} \, \ng \tilde\phi^3 f(1024 \tilde\phi^2)
           + {\textstyle {1\over 4!}} \, {N\lambda\over q^2} \, \tilde\phi^4
  \right] ,
\end {equation}
where $\tilde\phi \equiv M / (N q^2 \mu) = \phi/(N q \mu^{1/2})$.
The condition for a
first-order phase transition is that there exist a mass $\tilde m$ and
field $\tilde\phi \not= 0$ such that $V(\tilde \phi) = V'(\tilde\phi) = 0$.
By applying
$(1 - {\textstyle {1\over2}} (\tilde\phi \partial/\partial \tilde\phi))$
to Eq.~(\ref {eqnv}), one may derive the constraint
\begin {equation}
\label {eqnfconst}
   {\partial \over \partial \tilde\phi}
    \left( \tilde\phi f(1024 \tilde\phi^2) \right)
    = {1\over4\ng} {N\lambda\over q^2} \, \tilde\phi \,.
\end {equation}
The left-hand side is plotted in Fig.~\ref {figab}, and the right-hand side is
depicted by the dashed line.
Because the left-hand side is concave downward,
it is clear that the maximum value of $N\lambda/q^2$ which
satisfies the constraint is
\begin {equation}
\label {eqnlim}
	{N\lambda\over q^2}
    =
	4\ng \left({\partial \over \partial \tilde\phi} \right)^2
	\left[ \tilde\phi f(1024 \tilde\phi^2) \right]
	\biggr|_{\tilde\phi=0}
    =
	{96 \, \ng \over \pi^2} .
\end {equation}
This is precisely the condition that the
coefficient of $\tilde \phi^4$ vanish in the expansion of $V(\tilde \phi)$
about $\tilde \phi = 0$, which is the standard mean-field criterion for
a tricritical point.

Before comparing this result to the $\eps$-expansion result of
(\ref{eqxcr}), one must convert the definitions of $\lambda$ and
$q^2$, which are scheme-dependent.  In our three-dimensional analysis,
$\lambda$ and $q^2$ have simply referred to the (dimensionless) bare couplings
$\lambda_0$ and $q_0^2$,
which do not require any ultraviolet renormalization.
In the $\eps$-expansion, however, our $\lambda$ and $q^2$ were
defined by minimal subtraction in four dimensions, so that
\begin{eqnarray}
   \lambda_0 &=&
   \lambda + {1\over\eps} (a \, q^4 + b \, q^2 \lambda + c \, \lambda^2)
                       + O(q^6) ,
\\
   q_0^2    &=& q^2 + {1\over\eps} \beta_0 \, q^4 + O(q^6) ,
\end{eqnarray}
Amusingly,
at leading order in $1/N$,
$(\lambda_0,q_0^2)$ at the tricritical point
is twice $(\lambda,q^2)$ and so the ratio $\lambda/q^2$ is not changed
by the redefinition.

The three-dimensional result (\ref{eqnlim})
has the same $N$ and $n_g$ dependence as does the $\epsilon$-expansion
result (\ref{eqxcr}), but the coefficient differs from the first term
of the $\eps$-expansion by a factor of roughly five.
As discussed before, the ill-behaved $\eps$-expansion (\ref{eqxcr})
clearly cannot be used to obtain a meaningful estimate of
the overall coefficient at $\epsilon = 1$.

The preceding discussion of the scheme-dependence of $\lambda$
and $q^2$ brings up an important point.
The ratio $\lambda/q^2$ at the tricritical point is {\it not\/}
a physical quantity; it is scheme-dependent and will change under
generic finite renormalizations.  At the very end of section
\ref{2 loop latent sec} we discussed
the dangers of testing the $\eps$-expansion with unphysical
quantities.
One should really test the $\eps$-expansion in the case at hand
with something physical.
A good choice would to consider the RG flow line that connects the
Gaussian fixed point at $\lambda=q^2=0$ with the tricritical point.
The slope $\lambda/q^2$ of the line {\it at} the Gaussian fixed point
is invariant under finite, perturbative redefinitions of the couplings
because the couplings are arbitrarily small near the Gaussian
fixed point.
Within the $\eps$-expansion, the leading-order result for this
slope is no different from the result $54 \, \ng/N$
at (\ref{eqxcr}) for the tricritical point, and so again
differs from the three-dimensional result by a factor of
roughly five.  We have not computed the expansion of this
physical ratio $\lambda/q^2$ to higher orders in $\eps$.

Contrary to our results here,
Jain and Papadopoulos saw no evidence of a first-order
phase transition for {\it any\/} value of $\lambda/q^2$ of order $1/N$.
They missed it because they did not look for the transition in the
proper range of $\phi$ and $m^2$, as given by Eq.~(\ref {eqnord}).

\section {Conclusions}

We have applied $\eps$-expansion techniques to the study of
first-order electroweak phase transitions.
The $\eps$-expansion is not always reliable, but it provides
its own diagnostic for failure.
A comparison of whether sub-leading terms in the $\eps$-expansion
are larger or smaller than the leading term provides an
{\it a posteriori\/} indication of whether the $\eps$-expansion
is likely to be a reasonable approximation for the problem at hand.

The $\eps$-expansion has had some spectacular successes, such as
the computation of critical exponents in the Ising model.
Assessing its utility in gauged Higgs theories is more complicated.
By comparing with large $N$ calculations, we have found that
the $\eps$-expansion works qualitatively but
does not provide quantitatively useful predictions at $\eps = 1$
when the number of scalar fields is large.
Nevertheless,
in this case, we reassuringly also found that
the sub-leading terms in the $\eps$-expansion were large compared
to the leading one; so the expansion diagnosed its own failure.

In contrast, our results for real problems of interest
at small $N$ are much more encouraging.
Several tests of the $\eps$-expansion in the perturbative
regime $\lambda \ll g^2$, corresponding to very light Higgs masses,
showed that the expansion is quite well-behaved.
However, what we really want to know is how well the
$\eps$-expansion does when $\lambda \gsim g^2$ and conventional
perturbation theory fails.
We have made one non-trivial test in this case by computing to next-to-leading
order in $\eps$ 
the latent heat of the transition
(scaled by the correlation length squared).
For Higgs masses below 100--200 GeV, this $\eps$-expansion
does not seem too bad.  In contrast, we know that for some
quantities conventional perturbation theory fails badly for Higgs masses
as small as 35 GeV.
We have only made one such test of the $\eps$-expansion
for $\lambda \gsim g^2$, however, and more are certainly called for.
It's always possible that we happened to pick a particular quantity that
is better behaved than most.

If, based on the above evidence, we tentatively
assume that the $\eps$-expansion is a reasonable
approximation, then
we can glean insight from a number of
relatively simple {\it leading}-order calculations.
In particular, leading-order calculations predict that the
transition is stronger than that given by conventional
one-loop (ring-improved) perturbation theory when the Higgs mass
is below roughly 130 GeV.
However,
in contrast to intuition based on the {\it tree}-level
relation between the sphaleron mass and the Higgs expectation
$\vev$, the rate of baryon number non-conservation in the asymmetric
phase is always
{\it larger}.
Even if, in the end, one finds that these calculations
do not extrapolate well from $4{-}\eps$ to three dimensions, there
is still an important lesson to extract.
If the equilibrium properties of the electroweak phase
transition are studied on a lattice \cite {Kajantie},
and if the phase transition
is found to be considerably stronger than expected, this does not
necessarily make electroweak baryogenesis more viable in the
model studied; an unexpectedly strong transition is not necessarily
one with an correspondingly small rate of baryon nonconservation at
the completion of that transition.

Based on the success of our main test of the $\eps$-expansion,
it is possible that the expansion will provide a useful
quantitative as well as qualitative tool for cases of interest
to electroweak baryogenesis.
The particular application to the baryon nonconservation rate,
however, forces one into awkward technical contortions
related to the intrinsically three-dimensional nature of
the sphaleron.
More study will be needed to see if the suggestions we have
made will really produce a successful method for computing
this rate to next-to-leading order in the $\eps$-expansion.

\bigskip

This work was supported by the U.S. Department of Energy,
grant DE-FG06-91ER40614.
We thank Sergei Khlebnikov and Larry Sorensen for useful discussions.

\newpage
\appendix

\section {The two-loop potential}
\label {potential appendix}

The two-loop potential can be extracted from Ref.~\cite {Ford}
if one specializes
to U(1) or SU(2), and changes the overall group factor for each diagram
to generalize to an arbitrary number $N$ of complex scalar degrees of freedom.
Setting $\lambda$ to zero and ignoring $m^2$ compared to $M^2$, the
relevant terms in eqs.~(5.4-5) of Ref.~\cite {Ford} become
\begin {eqnarray}
    \mu^\epsilon V^{(2)} &=&
    \mu^\epsilon V^{(1)} + V_{\rm SV} + V_{\rm V} \,,
\\
\noalign {\hbox {where}}
  V_{\rm SV} &=& \ng q^2
      \{ {\textstyle {1\over2}} N A(0,0,z) - z \, B(z,z,0)
      \} \,,
\\
  V_{\rm V}^{\rm U(1)} &=& 0 \,,
\\
  V_{\rm V}^{\rm SU(2)} &=&
  q^2 \{ -2\Delta(z,z,z) + 6\Sigma(z,z) - 3A(0,0,z) \} \,.
\end {eqnarray}
The functions
$A$, $B$, $\Delta$ and $\Sigma$ are given in eqs.~(5.8-15) of
Ref.~\cite {Ford} and are:
\begin {eqnarray}
  A(0,0,z) &=& - z \, \hat I(z,0,0)
	       - {2 \, z \over 3} \bar J(z) ,
\\
  B(z,z,0) &=& 3 \hat I(z,z,0) - {1\over2} \hat I(z,0,0)
               - 2 \, \bar\eps I(z,z,0)
               + {1\over 2 \, z} \hat J(z,z)
	       + 3 \bar J(z) \,,
\\
  \Delta(z,z,z) &=& -{99\over4} z \hat I(z,z,z)
                    + {3\over4} z \hat I(z,0,0)
                    + 18 \, z \bar\eps I(z,z,z)
\nonumber \\ && {}
                    + 9 \, \hat J(z,z)
                    - {3\over2} \left({1\over4 {-} 2\bar\eps}+4\right)
                          \bar\eps J(z,z)
                    - 61 \, z \bar J(z) \,,
\\
  \Sigma(z,z) &=& {27\over4} \hat J(z,z)
                  + \left( {(3 {-} 2\bar\eps)^3 \over 4 {-} 2\bar\eps}
                                              - {27\over4} \right) J(z,z)
                  - 9 \, z \bar J(z) ,
\end {eqnarray}
where
\begin {equation}
    z \equiv q^2 \bar\phi^2
\end {equation}
and we have defined $\bar J(x)$ in terms of their $J(x)$ by%
\footnote
    {%
    There are typographic or notational errors in Ref.~\cite {Ford};
    equations (5.8-15) are correct only if each explicit factor
    of $J(w)$ in those equations (but not $J(v,w)$ or $\hat J(v,w)$)
    is replaced by $\bar J(w)$.  Also, the factors of $\mu^{2\bar\eps}$
    in their Eqs.~(3.4-5) should be eliminated.
    }
\begin {equation}
    \bar J(x) = J(x) / (4 \pi)^2 \,.
\end {equation}
Their definition of $\eps$, written as $\bar\eps$ in this appendix, is
related to ours by
\begin {eqnarray}
   \bar\eps &=& \eps/2 \,.
\end {eqnarray}
The functions $J$, $\hat J$, $I$, and $\hat I$
are defined in Eqs.~(3.1-5) of Ref.~\cite {Ford}.
The special cases of these functions which
are needed may be extracted from eqs.~(4.13),
(4.16-7) and (4.19) of Ref.~\cite {Ford} and are:
\begin {eqnarray}%
 \bar J(z) &=& {z \over (4\pi)^4}
		\left({ z \over 4\pi \mu^2}\right)^{-\bar\eps}
		\Gamma (-1 {+} \bar\eps) \,,
\\
    J(z,z) &=& {z^2 \over (4\pi)^4}
		\left({ z \over 4\pi \mu^2}\right)^{-2\bar\eps}
		\Gamma (-1 {+} \bar\eps)^2 \,,
\\
  I(z,0,0) &=& {z \over (4\pi)^4}
		\left({ z \over 4 \pi \mu^2}\right)^{-2\bar\eps}
               {\Gamma(\bar\eps) \Gamma(-1 {+} 2\bar\eps) \Gamma(1 {-}
\bar\eps)
                  \over 1-\bar\eps} \,,
\\
  I(z,z,0) &=& {z \over (4\pi)^4}
		\left({ z \over 4 \pi \mu^2}\right)^{-2\bar\eps}
               {\Gamma(\bar\eps) \Gamma(-1 {+} \bar\eps) \over 1-2\bar\eps} \,,
\\
  I(z,z,z) &=&
               {z \over (4\pi)^4}
	         \left({ z \over 4 \pi \mu^2}\right)^{-2\bar\eps}
		 {\textstyle {3\over2}} \, r(2\bar\eps) \,
                 \Gamma(\bar\eps) \Gamma(-1 {+} \bar\eps)
	      + I(\sqrt{3}z,0,0) \sin(\bar\eps\pi) \,,
\\
   \hat J(z,z) &=& J(z,z) + 2 \eps^{-1} z \bar J(z) \strut \,,
\\
   \hat I(z,0,0) &=& I(z,0,0) - \eps^{-1} \bar J(z) \,,
\\
   \hat I(z,z,0) &=& I(z,z,0) - 2 \eps^{-1} \bar J(z) \,,
\\
   \hat I(z,z,z) &=& I(z,z,z) - 3 \eps^{-1} \bar J(z) \,.
\end {eqnarray}%
We have also used the limiting values
$
   J(0) = J(z,0) = \hat J(z,0) = C(z,0) = 0
$
to simplify the results of Ref.~\cite {Ford}.
The function $r$ appearing in $I(z,z,z)$ is
\begin {equation}
   r(2\bar\eps)
   = 2^{2\bar\eps} \int\nolimits_0^1 {\d t\over(t^2+3)^{\bar\eps}}
   = \left(4\over3\right)^{\bar\eps} \,
        {}_2 F_1 \left(\bar\eps,{1\over2};{3\over2};-{1\over3}\right)
\label {rdef}
\end {equation}
and has the limits
\begin {eqnarray}
   r(x) &=& 1 + x \left[ 1 - {\sqrt 3 \pi\over 6} \right]
              + x^2 \left[ 1 + \sqrt{3} L\left(\pi\over6\right)
		 - {\sqrt{3} \pi\over 6} (1 + \ln2 - {\textstyle {1\over2}}\ln3)
                \right]
            + O(x^3) \,,
\\
   r(1) &=& 2 \, {\rm Sinh}^{-1} (1/\sqrt{3}) \,.
\end {eqnarray}
Here, $L(t)$ is Lobachevskiy's function, defined in Eq.~(\ref {lobachevsky}).
Note that to get a result for general $\bar\eps$,
it is important to use formulas (3.3) and (4.19) of Ref.~\cite {Ford}
where the limit $\bar\eps \rightarrow 0$ has not yet been taken.

Putting everything together, and switching from
$\bar\eps$ to our canonical definition of $\eps$,
gives the following results for U(1) and SU(2),
\begin {eqnarray}
   \mu^\eps V^{(2)}_{\rm U(1)} &=&
   \mu^\eps V^{(1)}_{\rm U(1)} +
   {q^2 \mu^4 \over(4\pi)^2}
	\Biggl\{
	    \Biggl[
	       - {3-\eps \over 1-\eps} \,
		    \Gamma(-1 {+} {\textstyle {1\over2}} \eps) \,
		    \Gamma({\textstyle {1\over2}} \eps)
	       - {\textstyle {1\over2}}
		    \Gamma(-1 {+} {\textstyle {1\over2}} \eps)^2
\nonumber \\ && \qquad \qquad\qquad\qquad {}
	       + {1-N \over 2-\eps} \,
		    \Gamma(1 {-} {\textstyle {1\over2}}\eps) \,
		    \Gamma({\textstyle {1\over2}}\eps) \,
		    \Gamma(-1 {+} \eps)
	    \Biggr]
		\left( q^2\bar\phi^2\over4\pi\mu^2 \right)^{2-\eps}
\nonumber \\ && \qquad\qquad\qquad {}
		+ (9+N) \left( {1\over\eps} - {1\over3} \right)
		    \Gamma(-1 {+} {\textstyle {1\over2}} \eps)
		\left( q^2\bar\phi^2\over4\pi \mu^2 \right)^{2-\eps/2}
\nonumber \\ && \qquad\qquad\qquad {}
		+ \left[ {9+N\over\eps^2} - {45+7N \over 6\eps} \right]
		\left( q^2\bar\phi^2\over4\pi\mu^2 \right)^2
       \Biggr\}
       \times \left[1 + O\left(m^2\over M^2\right) \right] ,
\end {eqnarray}
\begin {eqnarray}
   \mu^\eps V^{(2)}_{\rm SU(2)} &=&
   \mu^\eps V^{(1)}_{\rm SU(2)} +
   {3q^2 \mu^4 \over(4\pi)^2}
   \Biggl\{
	\Biggl[
		\left(  - {3 - \eps \over 1-\eps}
			+ {9\over4} \, (11-4\eps) \, r(\eps) \right)
		\Gamma(-1 {+} {\textstyle {1\over2}} \eps ) \,
		\Gamma({\textstyle {1\over2}} \eps )
\nonumber \\ && \qquad\qquad\qquad\qquad {}
	   + {11-4\eps \over 2-\eps} \, 3^{(3-\eps)/2} \pi
                \Gamma(-1 {+} \eps)
           + (7 - 8\eps + 2\eps^2) \,
		\Gamma(-1 {+} {\textstyle {1\over2}} \eps)^2
\nonumber \\ && \qquad\qquad\qquad\qquad {}
           + {2-N \over 2-\eps} \,
                \Gamma(1 {-} {\textstyle {1\over2}} \eps) \,
                \Gamma({\textstyle {1\over2}} \eps) \,
		\Gamma(-1 {+} \eps)
       \Biggr]
	    \left( {q^2\bar\phi^2 \over 4\pi \mu^2 }\right)^{2-\eps}
\nonumber \\ && \qquad\qquad\qquad {}
	   + (-61+N) \left( {1\over\eps}-{1\over3} \right)
		\Gamma(-1 {+} {\textstyle {1\over2}} \eps)
	    \left( {q^2\bar\phi^2 \over 4\pi \mu^2 }\right)^{2-\eps/2}
\nonumber \\ && \qquad\qquad\qquad {}
	   + \left[ {-61+N\over\eps^2} + {7 \, (73-N)\over 6\eps} \right]
	     \left({ q^2\bar\phi^2 \over 4\pi \mu^2 }\right)^2
    \Biggr\} \times \left[1 + O\left(m^2\over M^2\right) \right] .
\end {eqnarray}

\section {The two-loop scalar mass $\beta$-function}
\label {beta-m appendix}

Ref.~\cite {Ford} gives standard model results for the $\beta$-function
for the scalar mass.
One may easily generalize these results to the case of an arbitrary number
of scalar fields, just as was done for the two-loop potential in
Appendix~\ref{potential appendix}.
The only change necessary is to insert the appropriate $N$ dependence into
the group factor of each graph, which changes Eqs.~(5.2) and (5.4) of
Ref.~\cite {Ford} to
\begin {eqnarray}
  V_{\rm S} &=& {-\lambda^2\phi^2 \over 12}
                   \left[
		   \hat I(H,H,H) +
		   {\textstyle {1\over3}} (2N {-} 1) \, \hat I(H,G,G)
		   \right]
\nonumber
\\ && \!\!\quad{}
                + {\lambda\over8} \, \left[
                     \hat J(H,H) +
		     {\textstyle {2 \over 3}} (2N {-} 1) \, \hat J(H,G) +
		     {\textstyle {1 \over 3}} (4N^2 {-} 1) \, \hat J(G,G)
                  \right]^{\strut}_{\strut} ,
\\
  V_{\rm SV} &=& {\ng \, q^2 \over 2} \> \Bigl[
                     A(H,G,Z) + (N {-} 1) \, A(G,G,Z) - 2 Z \, B(Z,Z,H)
\nonumber
\\ && \quad\qquad {}
                      + C(Z,H) + (2N {-} 1) \, C(Z,G) \Bigr] ,
\\
\noalign {\hbox {where}}
    H &=& m^2 + {\lambda \over 2} \, \phi^2 \,, \qquad
    G = m^2 + {\lambda \over 6} \, \phi^2 \,, \qquad \hbox {and} \qquad
    Z = q^2 \, \phi^2 \,.
\end {eqnarray}
Following the derivation of Ref.~\cite {Ford}
and inserting the appropriate Landau gauge values of the
scalar anomalous dimension \cite {two-loop-beta},
\begin {eqnarray}
    \mu {d \phi \over d \mu} &=&
    - \left[ \gamma^{(1)} + \gamma^{(2)} + O(\lambda^3,q^6) \right]
\\
\noalign {\hbox {with}}
    (4\pi)^2 \, \gamma^{(1)} &=& -3 \, \ng \, q^2 \,,
\\
    (4\pi)^4 \, \gamma^{(2)}_{\rm U(1)}
    &=&
	{\textstyle {1 \over 18}} \, (N {+} 1) \, \lambda^2 +
	{\textstyle {1 \over 6}} \, (11N {+} 9) \, q^4 \,,
\\
    (4\pi)^4 \, \gamma^{(2)}_{\rm SU(2)}
    &=&
	{\textstyle {1 \over 18}} \, (N {+} 1) \, \lambda^2 +
	{\textstyle {1 \over 2}} \, (11N {-} 533) \, q^4 \,,
\end {eqnarray}
then yields the $\beta$-functions given by our (\ref{kcoeff}).
Because we are interested in non-zero values of $\lambda$,
evaluating the resulting expressions is
slightly more complicated than was the case for
Appendix~\ref{potential appendix};
however it is also easier since one may
restrict attention to the $\eps\to0$ limit.

\end {document}